\documentclass[a4paper,fleqn]{cas-dc}

\usepackage[numbers]{natbib}

\usepackage{amsmath,amssymb,amsfonts}
\usepackage{algorithmic}
\usepackage{graphicx}
\usepackage{textcomp}
\usepackage{xcolor}
\usepackage[justification=centering]{caption} %
\usepackage{subcaption}
\usepackage{mathtools}
\usepackage{bm}
\usepackage{diagbox}
\usepackage{multirow}
\usepackage{makecell}
\usepackage{tabularx}

\begin{document}
\let\WriteBookmarks\relax
\def\floatpagepagefraction{1}
\def\textpagefraction{.001}
\shorttitle{}
\shortauthors{Y. Li et al}

\title [mode = title]{Spherical Pendulum with Quad-Rotor Thrust Vectoring Actuation — A Novel Mechatronics and Control Benchmark Platform}                      
   
\author[1]{Yuchen Li}
\ead{lyc1998kumamon@g.ucla.edu}
\credit{Writing – original draft and revision, Design-and-Build, Methodology, Validation, Analyses}

\affiliation[1]{organization={Department of Mechanical \& Aerospace Engineering, University of California Los Angeles},
addressline={420 Westwood Plaza}, 
city={Los Angeles},
postcode={90095}, 
state={California},
country={USA}}

\author[1]{Omar Curiel}
\ead{omarzcuriel14@g.ucla.edu}
\credit{Writing – review \& editing, Methodology, Validation}

\author[2]{Sheng-Fan Wen}
\ead{fan790722@gmail.com}
\credit{Writing – review \& editing, Methodology, Validation}

\affiliation[2]{organization={Mechanical and Mechatronics Systems Research Labs, Industrial Technology Research Institute},
addressline={Chung Hsing Rd., Chutung}, 
postcode={310401}, 
city={Hsinchu},
state={Taiwan},
country={R.O.C.}}

\author[1]{Tsu-Chin Tsao}
\cormark[1]
\ead{ttsao@g.ucla.edu}
\credit{Writing – original draft and revision,  Concept Design, Methodology, Analyses}

\cortext[cor1]{Corresponding author}
\nonumnote{S. Wen was a visiting researcher at the University of California}

\begin{abstract}
Motor-actuated pendulums have been established as arguably the most common laboratory prototypes used in control system education because of the relevance to robot manipulator control in industry. Meanwhile, multi-rotor drones like quadcopters have become popular in industrial applications but have not been broadly employed in control education laboratory. Platforms with pendulums and multi-rotor copters present classical yet intriguing multi-degree of freedom (DoF) dynamics and coordinate systems for the control system investigation. In this paper, we introduce a novel control platform in which a 2-DoF pendulum capable of azimuth and elevation rotation is actuated through vectored thrust generated by a quadcopter. Designed as a benchmark for mechatronics and nonlinear control education and research, the system integrates detailed mechatronic implementation with different control strategies. Specifically, we apply and compare small perturbation linearization (SPL), state feedback linearization (SFL), and partial feedback linearization (PFL) to the nonlinear system dynamics. The performances are evaluated by time specifications of step response and Root-Mean-Square (RMS) error of trajectory tracking. The robustness of the closed-loop system is validated under external disturbances, and both simulation and experimental results are presented to highlight the strengths and limitations of the nonlinear model-based control approaches.
\end{abstract}

\begin{keywords}
Spherical Pendulum \sep Quadcopter Thrust Vectoring \sep Mechatronics \sep Nonlinear Control \sep Control Systems
\end{keywords}
\maketitle
\section{Introduction}

Feedback control systems are the cornerstones of numerous technological advances and are ubiquitous in modern industrial practices. The interdisciplinary subject materials, which are taught in several engineering disciplines, are analytical and abstract. To relate to the real world, it has long been recognized and encouraged to develop control systems laboratory and make experimental projects as an integral part of control engineering education at both undergraduate and graduate levels \cite{b1}\cite{b2}\cite{b3}. Motorized pendulums in various forms are by far the most popular among the wealth of the experimental platforms served as benchmarks for both educational and research purposes \cite{b4}\cite{b5}. Meanwhile, multi-rotor aerial platforms, particularly quad-rotor copters (quadcopters), have received increasing attention \cite{b6}\cite{b7}\cite{b8}. The over-actuated platforms enabled by multi-rotors or thrust vectoring \cite{b16}\cite{b17}\cite{b18} provide a new aspect that pendulums are so far lacking. Ryll et al. \cite{b10} developed a comprehensive modeling for a quadrotor where each propeller can be tilted. Gerber \cite{b9} further proposed a mechanism to tilt and twist at the same time using separate servos for full thrust vectoring. Bodie et al. \cite{b11} presented an omnidirectional aerial vehicle that can exert a wrench in any direction with six rotatable arms. Su et al. \cite{b12}\cite{b13} replaced the propeller with small conventional quadcopters and gimbal mechanism to establish a configurable aerial platform. Notably, multi-rotor copters and pendulums share similar salient features as benchmark platforms for education and research: simple and cost-effective modular components that may have various configurations, clean mathematical models that may include a range of linear and nonlinear dynamic behaviors and associated control challenges, and perceptible visibility of the dynamic motions.

A motorized pendulum platform is mostly configured as one that is fully actuated where a pendulum is driven by a motor or single arm \cite{b30}, aka Inverted Pendulum, or under-actuated where a pendulum is driven by a motorized translational cart or multiple rotary links \cite{b31}, aka Furuta Pendulum. A quadcopter by its nature is under-actuated, where 6 Degree-of-Freedom (DoF) free space motions are actuated by four rotor thrusts. While quadcopters may bring refreshing appeals to motivate learning, challenges in the flight space, durability and safety make them less practically useful than the anchored and confined pendulums. Combining quadcopters and pendulums to realize modular configurations and create a new class of benchmarks for control engineering education and research is our main objective in this paper.


Existing educational and research platforms that integrate pendulums with aerial rotors employ a variety of control strategies tailored to their specific purposes. As educational testbeds, some systems use a mechanical pendulum with one propeller at its free end or twin propellers mounted on a balancing beam. More advanced configurations include quadcopters with suspended passive pendulums for swing suppression and aerial manipulators with pendulum-like articulated arms. To expand the dynamics and control scopes of these existing pendulum-copter platforms, we aimed at creating a platform that may be configured as an over-actuated system, in addition to the common fully-actuated or under-actuated configurations (Table \ref{pendcopter_comparison}).
\begin{table*}[t]
\caption{Comparison of Pendulum–quadcopter Platforms and Related Systems}\label{pendcopter_comparison}
\centering
\renewcommand{\arraystretch}{1.5}
\begin{tabular}{|p{4.5cm}|p{6cm}|p{5cm}|}
\hline
\textbf{Platform} & \textbf{Control Approach} & \textbf{Remarks} \\
\hline
1-DoF Pendulum with Propellers & PI control \cite{b14} and PID control \cite{b15} & Excludes the nonlinear dynamics from the quadcopter \\
\hline
Suspended Load from UAV & Model
Predictive Control \cite{b23}, Backstepping Technique \cite{b24} and Geometric Control \cite{b25}  & Focuses on passive stabilization without active control of pendulum \\
\hline
Aerial Manipulator Systems & Robust H$\infty$ Control \cite{b26}, Force-position
Control \cite{b27} and Backstepping Approach \cite{b28} & Focuses on manipulation but not dynamic pendulum regulation \\
\hline
\textbf{Spherical Pendulum with Gimbal Actuator (this work)} & LQI and PID in cascaded architecture & Introduces a testbed for nonlinear, multi-configurable system control \\
\hline
\end{tabular}
\end{table*}

\begin{figure}[htbp]
    \centering
    \begin{subfigure}[b]{0.23\textwidth}
        \centering
        \includegraphics[width=\textwidth,height=0.7\textwidth]{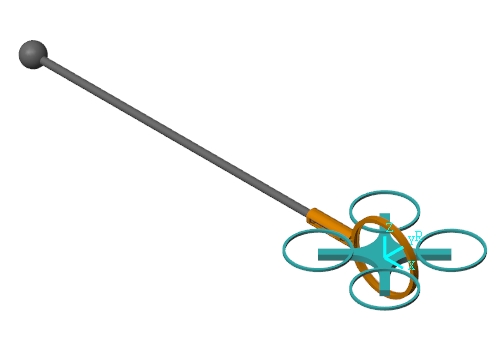}
        \caption{CAD Model} %
        \label{Setup_simulation}
    \end{subfigure}
    \begin{subfigure}[b]{0.23\textwidth}
        \centering
        \includegraphics[width=\textwidth,height=0.7\textwidth]{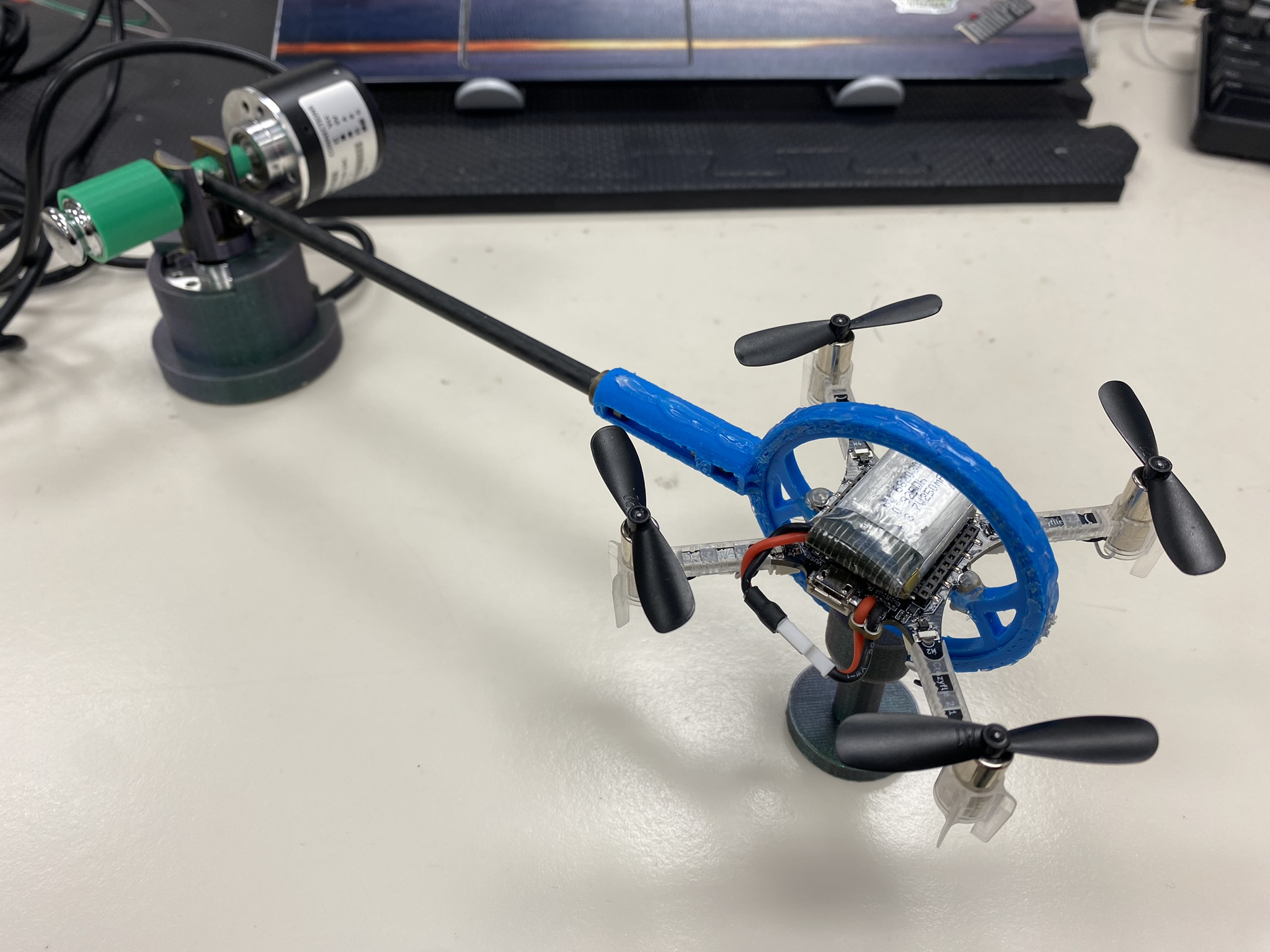}
        \caption{Real Plant} %
        \label{Setup_real_plant}
    \end{subfigure}
    \caption{Platform Configuration: (a) Computer Aided Design Model. (b) Realized Plant Assembly with Base Structure.}
    \label{Setup}
\end{figure}

Among various possible configurations, we created a 2-DoF pendulum with the azimuth and elevation angle and a 2-DoF gimbal, where a quadcopter is mounted at the center (Fig. \ref{Setup}). This spherical inverted pendulum can have several configurations: First, to simplify, the 2-DoF pendulum and gimbal can be mechanically constrained to become 1-DoF joint, where the motions are on a plane. Second, the four rotor thrusts are over-actuated for the 2-DoF pendulum but fully-actuated for the 4-DoF pendulum-copter. Third, the four rotor thrusts can be configured as a 3-DoF thrust vectoring actuator to be over-actuated for the 2-DoF pendulum but under-actuated for the 4-DoF pendulum-copter. This work mainly has three contributions to the literature: The platform presents rich scenarios of challenging dynamic and control problems with multi-configurable mechanical design. Furthermore, it demonstrates key control concepts and promotes visualization of stability and saturation, which motivates the innovation on equilibrium regulation, trajectory tracking and so forth. Finally, the proposed PFL method avoids the local limitation of SPL and the model fragility of SFL, focusing on the selected subset of the outputs to achieve the best performance. 

To the best of our knowledge, this is the first 4-DoF nonlinear pendulum dynamics and multi-rotor thrust vectoring platform, which may be formulated for all three actuation aspects. In this paper, we consider, among other possible configurations, the model-based control design \cite{b19} of the platform for the case of controlling the 2-DoF pendulum by the over-actuated 3-DoF quadcopter actuation. Three linearization methods: SPL, SFL and PFL are developed to control the angles of the pendulum and the gimbal. Moreover, the performance is evaluated by the time specifications of step responses for the gimbal and the Root-Mean-Square (RMS) error of the $\theta_1$ trajectory for the pendulum. The robustness of controllers are verified by adding external torque disturbances. The simulation and experiment results are both in time domains and they exhibit great effectiveness of controllers in terms of position control and simplifying coupled nonlinear systems.

The rest of this paper is organized as follows. The configuration setup and dynamics analysis are elaborated to generate the equations of motions in Section \ref{II}. The model-based hierarchical controller is designed for the pendulum and quadcopter respectively in Section \ref{III}. The gimbal control is verified by simulation and experiment for the step and ramp response in Section \ref{IV}. The pendulum control is implemented and compared to the simulation results for regulation and trajectory tracking cases in Section \ref{V}, followed by the conclusions given in Section \ref{VI}.

\section{Problem Formulation and Preliminaries}\label{II}

\begin{figure}[htbp]
    \centering
    \begin{subfigure}[b]{0.23\textwidth}
        \centering
        \includegraphics[width=\textwidth]{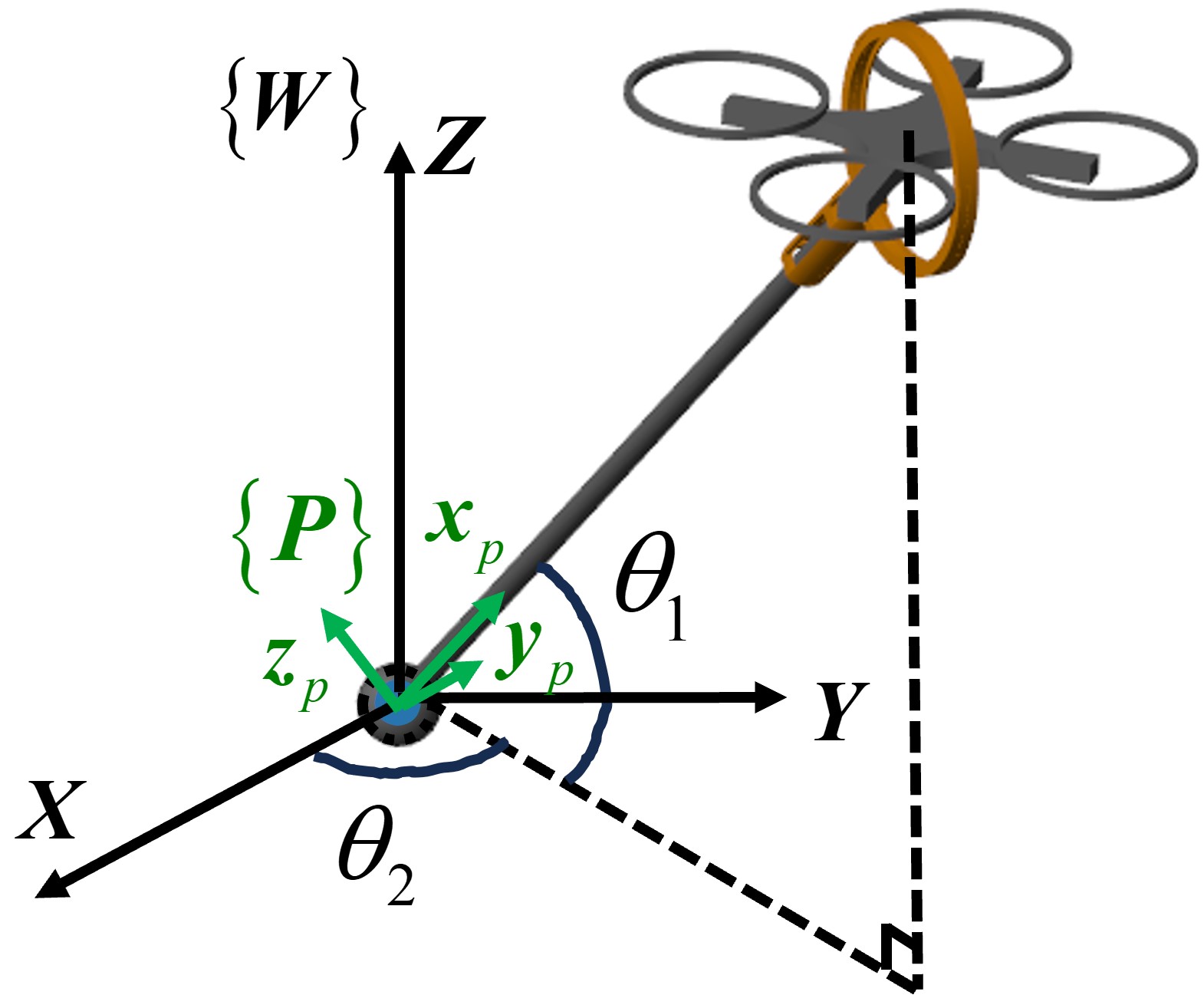}
        \caption{Isometric View}
    \end{subfigure}
    \begin{subfigure}[b]{0.23\textwidth}
        \centering
        \includegraphics[width=\textwidth]{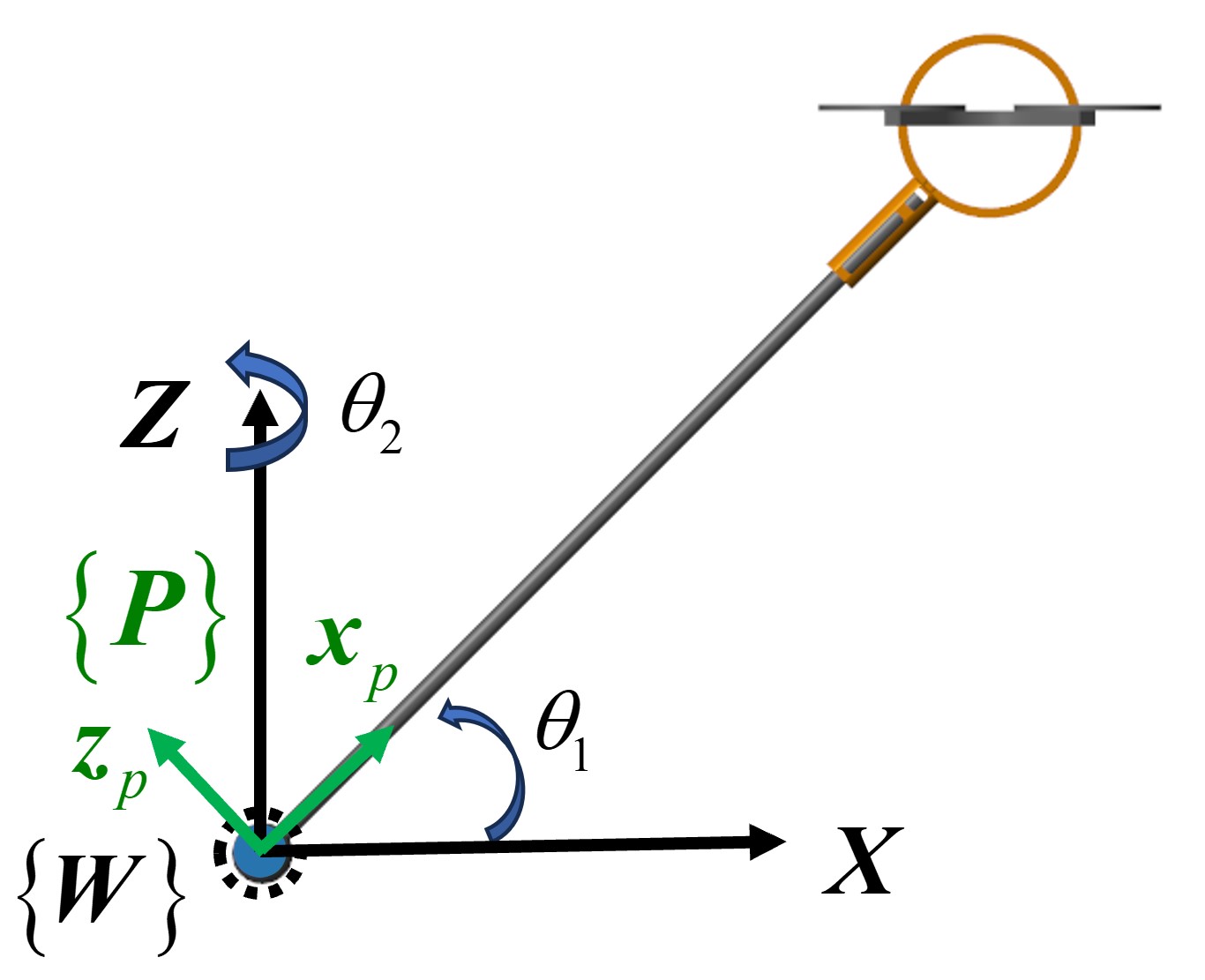}
        \caption{Side View}
        \label{pend_angles}
    \end{subfigure}
    \begin{subfigure}[b]{0.23\textwidth}
        \centering
        \includegraphics[width=\textwidth]{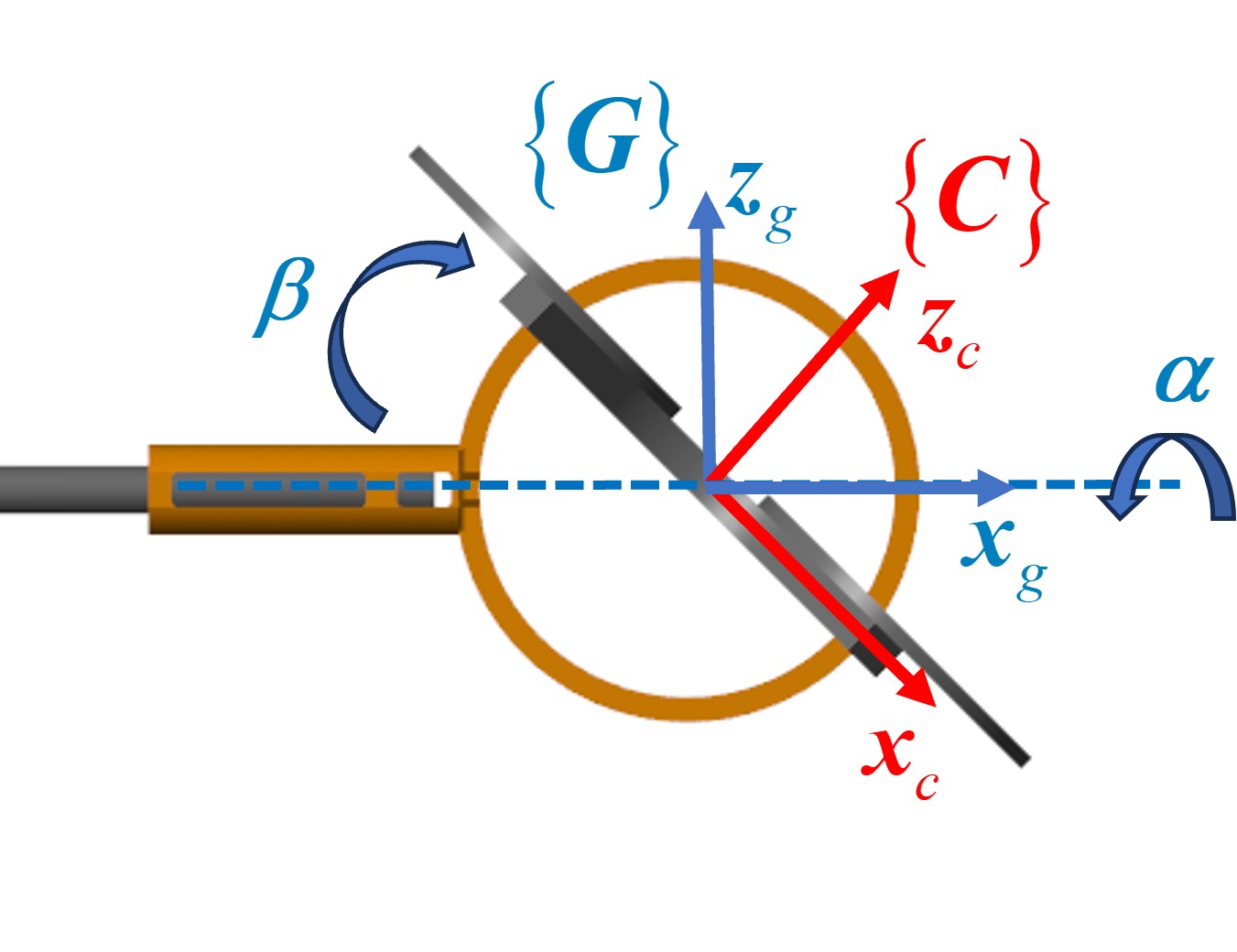}
        \caption{Gimbal View}
        \label{copter_angles}
    \end{subfigure}
    \caption{The Platform's Kinematics and Coordinate Frames: World Frame $\{W\}$, Pendulum Frame $\{P\}$, Gimbal Frame $\{G\}$ and Copter Frame $\{C\}$. Frame $\{P\}$ is rotated with the base while Frame $\{W\}$ is fixed. The x-axis in Frame $\{G\}$ is always aligned with the rod and Frame $\{C\}$ is attached on the board.}
    \label{Coordinates}
\end{figure}

The kinematic configuration of the proposed platform is defined by four coordinate frames attached to the components (Fig. \ref{Coordinates}). A joint rotates ($\theta_2$) with respect to a base fixed to the World Frame $\{W\}$'s +z axis. A second joint rotates ($\theta_1$) with respect to the first joint, the Pendulum Frame $\{P\}$'s -y axis. A 2-axis gimbal, which consists of an outer ring rotating ($\alpha$) with respect to the Pendulum Frame $\{P\}$'s +x axis and an inner ring rotating ($\beta$) with respect to the Gimbal Frame $\{G\}$'s +y axis. A quadcopter is mounted on the gimbal \cite{b20}, where the y-axis of the Copter Frame $\{C\}$ coincides with that of the Gimbal Frame $\{G\}$.

\subsection{System Dynamics}\label{II_A}
In the quadcopter, each propeller's rotation $\omega_i$ creates thrust and torque resisting the air $(i=1,2,3,4)$:
\begin{equation}
f_i = k_f\omega_i^2,\ \tau_i = k_m\omega_i^2
\end{equation}
 
\noindent
where $k_f$ and $k_m$ are coefficients depending on the air density and the rotor geometry. They create a resultant thrust and roll-pitch-yaw torques at the center:
\begin{equation}
\begin{bmatrix}
f \\ \tau_x \\ \tau_y \\ \tau_z
\end{bmatrix}=
\begin{bmatrix}
k_f & k_f & k_f & k_f \\ k_fd_r & k_fd_r & -k_fd_r & -k_fd_r \\
k_fd_r & -k_fd_r & -k_fd_r & k_fd_r \\ k_m & -k_m & k_m & -k_m
\end{bmatrix}
\begin{bmatrix}
\omega_1^2 \\ \omega_2^2 \\ \omega_3^2 \\ \omega_4^2
\end{bmatrix}
\end{equation}

\noindent
where $d_r$ is the distance between the center of the rotor and the $xy$ axes in Frame $\{C\}$. The quadcopter control input is therefore defined as $[f\ \bm{\tau}]^\mathrm{T}$, where $f$ stands for the thrust value provided and $\bm{\tau} \triangleq [\tau_x\ \tau_y\ \tau_z]^\mathrm{T}$ is the 3-DoF torque generated by the rotors.

Our platform consists of three components including a pendulum, a gimbal and a quadcopter, but the gimbal and the quadcopter are combined as one rigid body in this section. However, the simulation in Section \ref{IV} models the gimbal and the quadcopter as two rigid bodies. 

\begin{figure}[htbp]
    \centering
    \includegraphics[width=0.23\textwidth]{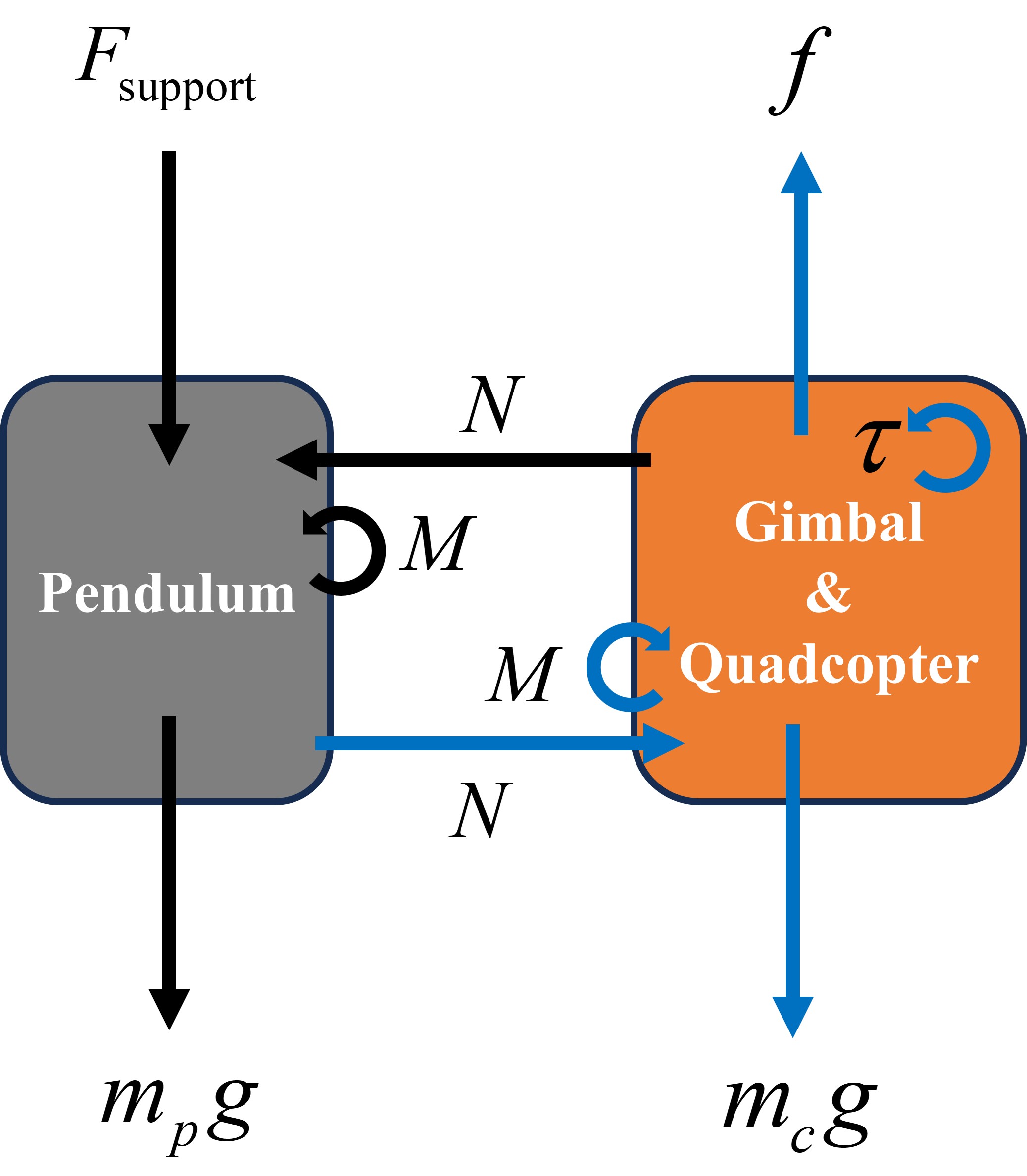}
    \caption{The Platform's Free-Body-Diagram: $m_p$ and $m_c$ are the mass of the pendulum and the quadcopter, respectively. $g$ is the constant gravitational acceleration. $F_{support}$ is the reaction force from the base fixed on the ground. $\bm{N}$ and $\bm{M}$ represent the interaction force and torque transmitted between the pendulum and the gimbal.}
    \label{FBD_system}
\end{figure}

According to the platform's Free-Body-Diagram (Fig. \ref{FBD_system}), the general Newton-Euler Equations can be applied:
\begin{equation}
\begin{array}{l}
    \bm{F_{net}} = m \bm{\dot{v}} \\
    \bm{M_{net}} = \bm{J\dot{\omega}} + \bm{\omega}\times(\bm{J\omega})
\end{array}
\label{New_Eul_general}
\end{equation}

\noindent
where $\bm{F_{net}}$ is the total force acting on the body and $\bm{M_{net}}$ is the total moment about the center of mass (CoM). $m$ is the mass and $\bm{J}$ is the inertia matrix with respect to the CoM of the body. $\bm{\dot{v}}$ and $\bm{\dot{\omega}}$ are the linear and angular accelerations.

For the gimbal-quadcopter rigid body:
\begin{align}
    m_c\bm{\dot{v}_c} = - &\prescript{W}{C}{\bm{R}}^\mathrm{T} m_cg \bm{\hat{z}}-\prescript{P}{C}{\bm{R}}^\mathrm{T} \prescript{P}{}{\bm{N}} + f \bm{\hat{z}} \label{gimbal_New} \\
    \bm{J_c}\bm{\dot{\omega}_c} + \bm{\omega_c} \times \bm{J_c}\bm{\omega_c} &= \bm{\tau} -\bm{\delta} \times \prescript{W}{C}{\bm{R}}^\mathrm{T} m_cg \bm{\hat{z}} \label{gimbal_Eul} \\ &-\prescript{P}{C}{\bm{R}}^\mathrm{T} \prescript{P}{}{\bm{M}} - \bm{M_f(\omega_c,\dot{\omega}_c)} \notag
\end{align}

\noindent
where $\bm{J_c}$ is the inertia matrix of the quadcopter. Its angular and linear velocities are represented by $\bm{\omega_c}$ and $\bm{v_c}$. $\prescript{X}{Y}{\bm{R}}$ is a rotation matrix from Frame $\{X\}$ to Frame $\{Y\}$ in Euclidean Space. $\bm{\hat{z}}=[0\ 0\ 1]^\mathrm{T}$ is a unit vector showing the +z axis. $\bm{\delta}$ is the deviation vector that points to the CoM of the gimbal-quadcopter from its geometric center, which is neglected in our case. $\bm{M_f(\omega_c,\dot{\omega}_c)}$ includes the nonlinear friction caused by angular velocity and acceleration of the gimbal \cite{b29}, which is combined into the transmitted torque $\prescript{P}{}{\bm{M}}$. The superscript $\prescript{P}{}{[*]}$ shows the vector $[*]$ described in Frame $\{P\}$.

For the pendulum rigid body:
\begin{align}
\bm{J_p} \bm{\dot{\omega}_p} + \bm{\omega_p} \times \bm{J_p}\bm{\omega_p} &= \bm{d_g}\times\prescript{P}{}{\bm{N}}-\bm{M_f(\omega_p,\dot{\omega}_p)} \label{pend_Eul} \\
&+ \prescript{P}{}{\bm{M}} - \bm{d_p}\times\prescript{W}{P}{\bm{R}}^\mathrm{T}m_pg\bm{\hat{z}} \notag
\end{align}

\noindent
where $\bm{J_p}$ is the inertia matrix of the pendulum with respect to the fixed point at the base. $\bm{\omega_p}$ is the angular velocity of the pendulum. $\bm{d_g}=[L_g\ 0\ 0]^\mathrm{T}$ and $\bm{d_p}=[L_p\ 0\ 0]^\mathrm{T}$ are vectors pointing from the origin of Frame $\{P\}$ to that of the gimbal and the CoM of the pendulum, respectively. $\bm{M_f(\omega_p,\dot{\omega}_p)}$ contains the nonlinear friction from the pendulum motion, which is negligible since the manipulation is slow enough.

As Frame $\{P\}$ and Frame $\{G\}$ are aligned along the pendulum, $\prescript{P}{C}{\bm{R}}$ is equivalent to $\prescript{G}{C}{\bm{R}}$ which is derived by sequential rotations of $\alpha$ and $\beta$ based on the geometric characteristics (Fig. \ref{copter_angles}):
\begin{equation}
\prescript{P}{C}{\bm{R}} = \prescript{G}{C}{\bm{R}}\triangleq
\begin{bmatrix}
c\beta & 0 & s\beta \\
s\alpha s\beta & c\alpha & -s\alpha c\beta \\
-c\alpha s\beta & s\alpha & c\alpha c\beta
\end{bmatrix}
\label{R_GC}
\end{equation}

Similarly, $\prescript{W}{P}{\bm{R}}$ is derived by sequential rotations of $\theta_2$ and $\theta_1$ based on the geometric characteristics (Fig. \ref{pend_angles}):
\begin{equation}
\prescript{W}{P}{\bm{R}}\triangleq
\begin{bmatrix}
c\theta_1c\theta_2 & -s\theta_2 & -s\theta_1c\theta_2 \\
c\theta_1s\theta_2 & c\theta_2 & -s\theta_1s\theta_2 \\
s\theta_1 & 0 & c\theta_1
\end{bmatrix}
\label{R_WP}
\end{equation}

To relate the quadcopter control input $[f\ \tau_x\ \tau_y\ \tau_z]^\mathrm{T}$ to the pendulum output $[\theta_1\ \theta_2]^\mathrm{T}$, the intermediate variables $\alpha$ and $\beta$ are needed. 

In Eq. (\ref{gimbal_Eul}), $\bm{\omega_c}$ can be composed of the angular velocities $\dot{\alpha}$ and $\dot{\beta}$ in Frame $\{G\}$:
\begin{equation}
\bm{\omega_c} = 
\begin{bmatrix}
    \dot{\alpha}c\beta & \dot{\beta} & \dot{\alpha}s\beta
\end{bmatrix}^\mathrm{T}
\label{gimbal_omega}
\end{equation}

\noindent
where $s[*]$ and $c[*]$ are abbreviations for $sin[*]$ and $cos[*]$. The rotational dynamics of the gimbal-quadcopter can be rearranged as:
\begin{small}
\begin{equation}
\begin{bmatrix}
J_{cx}c\beta & 0 \\ 0 & J_{cy} \\ J_{cz}s\beta & 0
\end{bmatrix}
\begin{bmatrix}
\Ddot{\alpha} \\ \Ddot{\beta}
\end{bmatrix}
+
\begin{bmatrix}
(J_{cz}-J_{cx}-J_{cy})s\beta\dot{\alpha}\dot{\beta} \\
(J_{cx}-J_{cz})s\beta c\beta\dot{\alpha}^2 \\
(J_{cy}-J_{cx}+J_{cz})c\beta\dot{\alpha}\dot{\beta}
\end{bmatrix}
=
\begin{bmatrix}
   \tau_x \\ \tau_y \\ \tau_z
\end{bmatrix}
-
\prescript{P}{C}{\bm{R}}^\mathrm{T} \prescript{P}{}{\bm{M}}
\label{gimbal_alpbet}
\end{equation}
\end{small}

\noindent
where $J_{cx}$, $J_{cy}$ and $J_{cz}$ are diagonal elements of $\bm{J_c}$. $J_{cx} = J_{cy}$ and the off-diagonal elements are zeros due to symmetry. This is the relation between $[\alpha\ \beta]^\mathrm{T}$ dynamics and $[\tau_x\ \tau_y\ \tau_z]^\mathrm{T}$.

In Eq. (\ref{pend_Eul}), $\bm{\omega_p}$ is related to the angular velocities $\dot{\theta}_1$ and $\dot{\theta}_2$ in Frame $\{W\}$:
\begin{equation}
\bm{\omega_p} = 
\begin{bmatrix}
    \dot{\theta}_2s\theta_1 & -\dot{\theta}_1 & \dot{\theta}_2c\theta_1
\end{bmatrix}^\mathrm{T}
\label{pend_omega}
\end{equation}

\noindent
where $\omega_{px}$ is mechanically restricted to zero due to the hardware configuration. Therefore, $\prescript{P}{}{\bm{N}}$ can be expressed from Eq. (\ref{gimbal_New}) with $\bm{\omega_p}$:
\begin{equation}
\begin{array}{l}
    \prescript{P}{}{\bm{N}} = -m_c\prescript{P}{}{\bm{\dot{v}_c}} - \prescript{W}{P}{\bm{R}}^\mathrm{T}m_cg \bm{\hat{z}} + \prescript{P}{C}{\bm{R}}f \bm{\hat{z}} \\
    \prescript{P}{}{\bm{\dot{v}_c}}\triangleq\bm{\dot{\omega}_p}\times\bm{d_g}
\end{array}
\label{gimbal_Np}
\end{equation}

Substituting Eqs. (\ref{R_GC})(\ref{R_WP})(\ref{pend_omega})(\ref{gimbal_Np}) into Eq. (\ref{pend_Eul}) yields:
\begin{small}
\begin{align}
    J_{px}\dot{\omega}_{px} &= (J_{pz}-J_{py})\dot{\theta}_1\dot{\theta}_2c\theta_1 + \prescript{P}{}{\bm{M}_x} \notag\\
    (J_{py}+m_cL_g^2)\dot{\omega}_{py} &= (J_{pz}-J_{px})\dot{\theta}_2^2s\theta_1c\theta_1-fL_gc\alpha c\beta \notag\\ 
    &+(m_cL_g+m_pL_p)gc\theta_1 + \prescript{P}{}{\bm{M}_y} \notag\\
    (J_{pz}+m_cL_g^2)\dot{\omega}_{pz} &= (J_{py}-J_{px})\dot{\theta}_1\dot{\theta}_2s\theta_1-fL_gs\alpha c\beta + \prescript{P}{}{\bm{M}_z} 
\label{pend_theta12}
\end{align}    
\end{small}


\noindent
where $$\dot{\omega}_{px} = \Ddot{\theta}_2s\theta_1+\dot{\theta}_1\dot{\theta}_2c\theta_1,\ \dot{\omega}_{py} = -\Ddot{\theta}_1,\ 
\dot{\omega}_{pz} = \Ddot{\theta}_2c\theta_1-\dot{\theta}_1\dot{\theta}_2s\theta_1$$

\noindent
$J_{px}$, $J_{py}$ and $J_{pz}$ are diagonal elements of $\bm{J_p}$. $J_{py} = J_{pz}$ because of symmetry. $J_{px}=0$ due to mechanical constraint and off-diagonal elements are zeros. This is the relation between $[\theta_1\ \theta_2]^\mathrm{T}$ dynamics and $[f\ \alpha\ \beta]^\mathrm{T}$.

Therefore, Eqs. (\ref{gimbal_alpbet}) and (\ref{pend_theta12}) are related by $[\alpha\ \beta]^\mathrm{T}$ dynamics, which constructs the model between $[f\ \tau_x\ \tau_y\ \tau_z]^\mathrm{T}$ and $[\theta_1\ \theta_2]^\mathrm{T}$.

\subsection{Control-Oriented Model}\label{II_C}

Since $\alpha$ and $\beta$ have free rotations (Fig. \ref{copter_angles}), the torques cannot be transmitted to the pendulum, i.e. $\prescript{P}{}{M_x} = 0$ and $\prescript{C}{}{M_y} = 0$. Applying these constraints to $\prescript{C}{}{\bm{M}} = \prescript{P}{C}{\bm{R}}^\mathrm{T}\prescript{P}{}{\bm{M}}$, we have:
\begin{equation}
\begin{array}{l}
\prescript{C}{}{M_x} = s\beta(s\alpha\prescript{P}{}{M_y}-c\alpha\prescript{P}{}{M_z}) \\
\prescript{C}{}{M_y} = c\alpha\prescript{P}{}{M_y}+s\alpha\prescript{P}{}{M_z} = 0 \\
\prescript{C}{}{M_z} = -c\beta(s\alpha\prescript{P}{}{M_y}-c\alpha\prescript{P}{}{M_z})
\end{array}
\end{equation}

Therefore, Eq. (\ref{gimbal_alpbet}) can be rearranged, multiplying the first row by $c\beta$, the third row by $s\beta$, and adding them, to eliminate $\prescript{P}{}{\bm{M}}$:
\begin{equation}
\begin{array}{l}
\Ddot{\alpha} = -\frac{2(J_{cz}-J_{cx})s\beta c\beta\dot{\alpha}\dot{\beta}}{J_{cx} c^2\beta + J_{cz} s^2\beta}  + \frac{\tau_\alpha}{J_{cx} c^2\beta + J_{cz} s^2\beta}
\\
\Ddot{\beta} = \frac{(J_{cz}-J_{cx})s\beta c\beta\dot{\alpha}^2}{J_{cy}} + \frac{\tau_\beta}{J_{cy}}
\end{array}
\label{gimbal_alpbet_sim}
\end{equation}

\noindent
where the control inputs are defined by $\bm{u_g}\triangleq[\tau_\alpha\ \tau_\beta]^\mathrm{T}$:
\begin{equation}
\begin{array}{l}
\tau_\alpha = \tau_xc\beta+\tau_zs\beta \\
\tau_\beta = \tau_y
\end{array}
\label{tau_ab}
\end{equation}

This shows that the gimbal dynamics and control are independent of those of the pendulum. To simplify the dynamics and control for the pendulum, it is desirable that the gimbal exerts the thrust vectoring force via $\bm{N}$ and negligible moment $\bm{M}$. To this end, we exploit the actuator redundancy and impose the third constraint on $\bm{M}$ by zeroing the moment along the +z axis of the gimbal by the actuator torque in the static sense, i.e. when the dynamic terms of Eq. (10) are neglected without affecting the gimbal dynamics:
\begin{equation}
    \prescript{G}{}{\tau_z} = \tau_zc\beta - \tau_xs\beta
\label{tau_gz}
\end{equation}

Therefore, Eqs. \ref{tau_ab} and \ref{tau_gz} uniquely determine the actuator torque:
\begin{equation}
\begin{array}{l}
    \tau_x = \tau_\alpha c\beta \\
    \tau_y = \tau_\beta \\
    \tau_z = \tau_\alpha s\beta
\end{array}
\label{tau_ab_input}
\end{equation}

\begin{figure}[htbp]
\centerline{\includegraphics[scale=0.28]{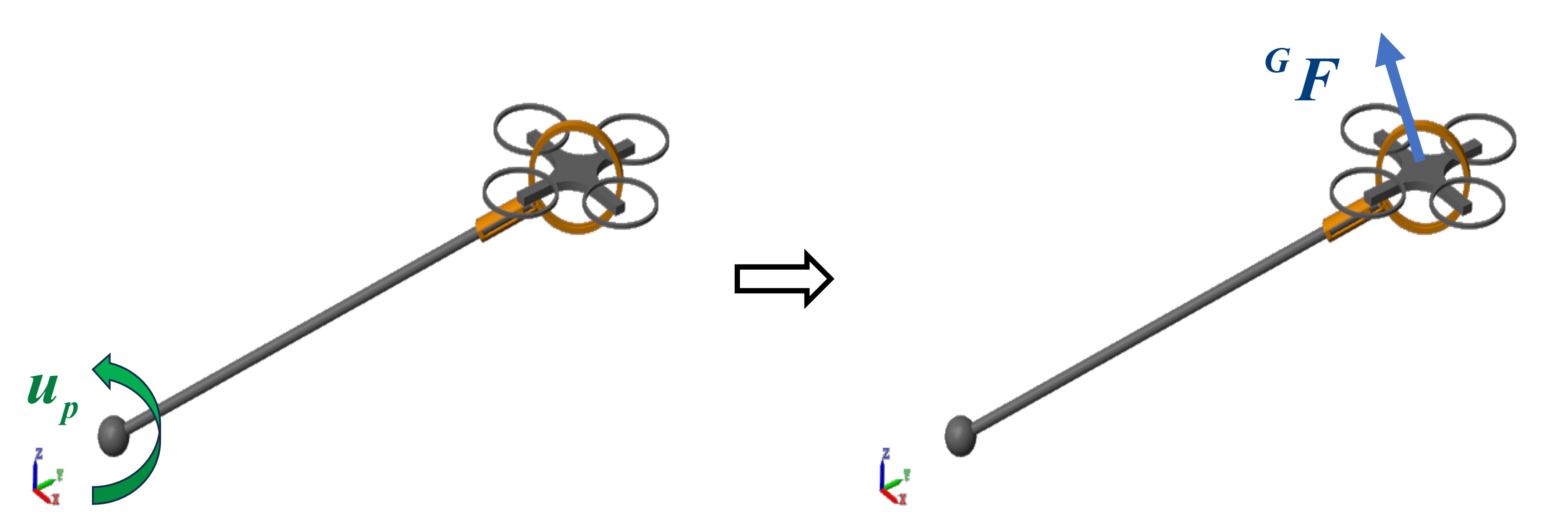}}
\caption{The mapping from the resultant torque in Frame $\{P\}$ to the desired thrust vector in Frame $\{G\}$.}
\label{geometric mapping relation}
\end{figure}

The control input of the pendulum system in Frame $\{P\}$ is defined by $\bm{u_p}\triangleq [T_x\ T_y\ T_z]^\mathrm{T}$, where $T_x$, $T_y$ and $T_z$ are resultant torques caused by an intermediate thrust vector in Frame $\{G\}$ (Fig. \ref{geometric mapping relation}). 

According to the coordinate transform, $\prescript{G}{}{\bm{F}} \triangleq[F_x\ F_y\ F_z]^\mathrm{T}$ is related to $f$ where $F_x$ can be any value due to redundancy:
\begin{equation}
    \prescript{G}{}{\bm{F}}=
    \prescript{G}{C}{\bm{R}}f\bm{\hat{z}}
    \Rightarrow
    \begin{array}{l}
    F_x = fs\beta \\
    F_y = -fs\alpha c\beta \\
    F_z = fc\alpha c\beta
    \end{array}
    \label{F_G}
\end{equation}

\noindent
where the inverse kinematics yields the mapping between $\prescript{G}{}{\bm{F}}$ and the intermediate variables $\triangleq[f\ \alpha\ \beta]^\mathrm{T}$ from Eq. (\ref{F_G}):
\begin{equation}
\begin{array}{l}
f = ||\bm{F}|| \\ \alpha = atan2(-F_y,F_z) \\ \beta = asin(F_x/f)
\end{array}
\label{f_alp_bet}
\end{equation}

Based on the torque definition in Frame $\{P\}$, $\bm{u_p}$ is related to $[f\ \alpha\ \beta]^\mathrm{T}$ by:
\begin{equation}
\bm{u_p} = \bm{d_g}\times\prescript{G}{}{\bm{F}}\Rightarrow
\begin{array}{l}
T_x = 0 \\
T_y = -F_zL_g = -fL_gc\alpha c\beta \\
T_z = F_yL_g = -fL_gs\alpha c\beta
\end{array}
\label{pend_torque}    
\end{equation}

Therefore, the rotational dynamics in Eq. (\ref{pend_theta12}) can be further simplified using $T_y$ and $T_z$:
\begin{equation}
\begin{array}{l}
    \Ddot{\theta}_1 = -\frac{(m_cL_g+m_pL_p)gc\theta_1+(J_{pz}-J_{px})\dot{\theta}_2^2s\theta_1c\theta_1}{J_{py}+m_cL_g^2} - \frac{T_y}{J_{py}+m_cL_g^2} \\
    \Ddot{\theta}_2 = \dot{\theta}_1\dot{\theta}_2tan\theta_1 + \frac{(J_{py}-J_{px})\dot{\theta}_1\dot{\theta}_2tan\theta_1}{J_{pz}+m_cL_g^2} + \frac{T_z/c\theta_1}{J_{pz}+m_cL_g^2}
\end{array}
\label{pend_theta12_sim}
\end{equation}

In summary, the whole system consisting of the pendulum, the gimbal and the quadcopter is combined together:
\begin{figure}[htbp]
\centerline{\includegraphics[scale=0.2]{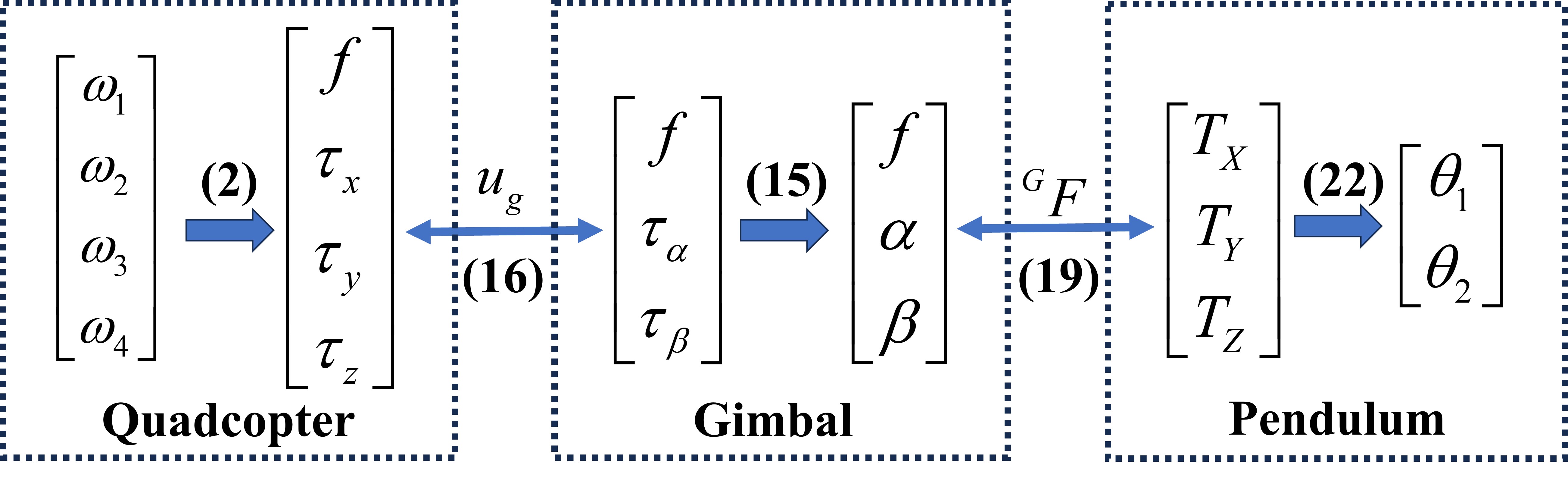}}
\caption{The whole system overview: The pendulum input is related to the gimbal output with an intermediate thrust vector $\prescript{G}{}{\bm{F}}$ where $F_x$ is determined by an Over-Actuation Mapping covered in Section \ref{III_C}. The gimbal input is fulfilled by four rotors using the conventional mixed algorithm for a quadcopter.}
\label{whole_system_overview}
\end{figure}

\section{Hierarchical Control}\label{III}
\begin{figure*}[htbp]
\centerline{\includegraphics[scale=0.3]{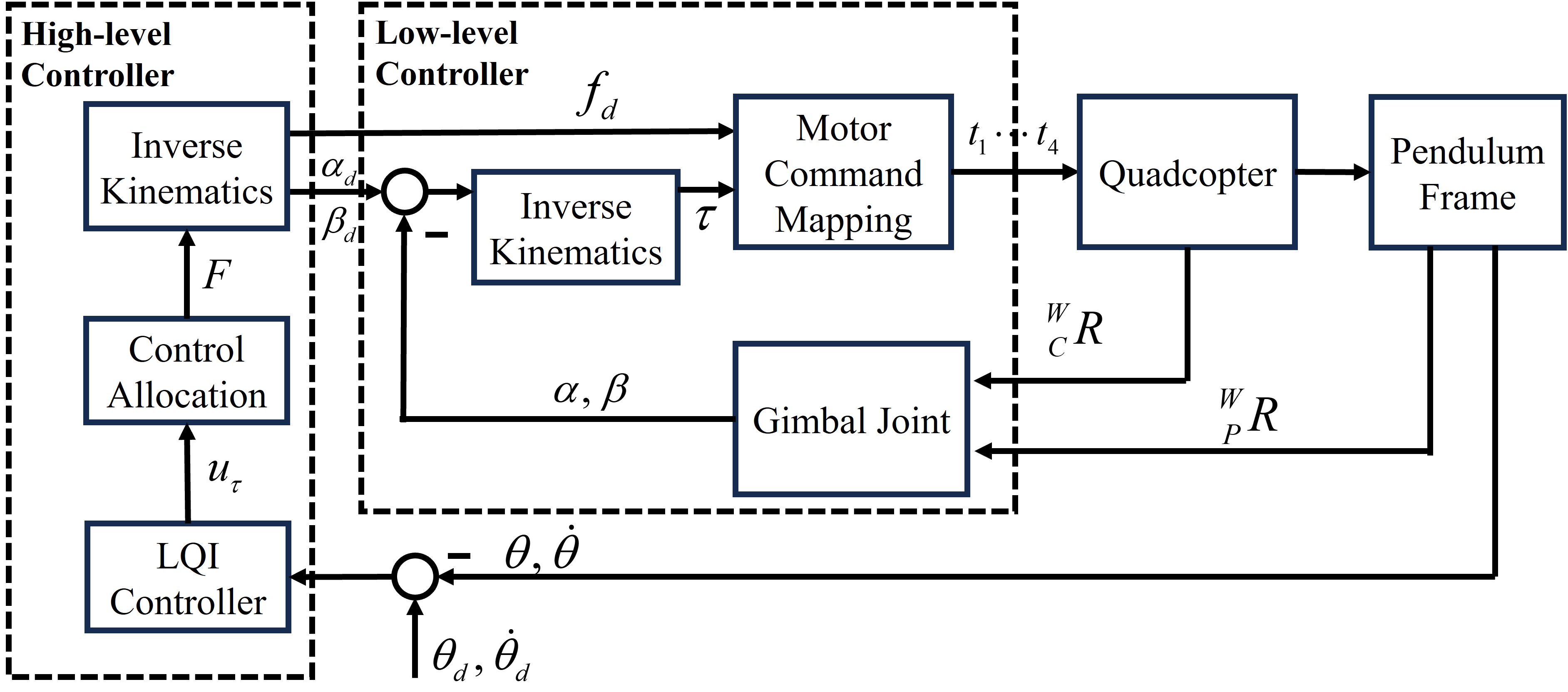}}
\caption{Block diagram of the 4-DoF platform. The high-level controller consists of a LQI controller to generate the input $\bm{u}_\tau$ to regulate $\theta_1$ and $\theta_2$. Then the equivalent torque command is mapped to a required thrust vector in Frame $\{G\}$ parameterized with $f$, $\alpha$ and $\beta$. The low-level controller consists of an on-board model-based PID controller designed to robustly track the thrust direction while meeting the thrust command via a feed-forward approach.}
\label{High_level_diagram}
\end{figure*}

 Hierarchical control, similar to the computed torque method in robot manipulator control, computes the desired torque on the pendulum and allocates the corresponding thrust vector command for the gimbal (Fig. \ref{High_level_diagram}). Such mapping is defined as an equivalent pendulum torque command \cite{b21} to be met by the gimbal vectoring thrust actuated by the quadcopter.

The continuous-time dynamics of both the pendulum and the gimbal are in the state-space form:
\begin{equation}
    \dot{\bm{x}}=\bm{F(x)}+\bm{G(x)u}
    \label{state_space}
\end{equation}

\noindent
where $\bm{x}$ is a state vector and $\bm{u}$ is a control input. It is nonlinear and highly coupled for the controller design, so linearization has to be done by introducing 
an equivalent pseudo-linear form:
\begin{equation}
\bm{\dot{x}}\triangleq\bm{Ax}+\bm{B\eta}
\label{pseudo_linear_form}
\end{equation}

\noindent
where $\bm{\eta}$ is an equivalent virtual input. The matrices $\bm{A}$ and $\bm{B}$ are time-invariant while $\bm{\eta}$ absorbs the nonlinear components. Therefore, a Linear Quadratic Integral (LQI) controller is proposed based on Eq. (\ref{pseudo_linear_form}).

In this Section, we consider three linearization methods. Small Perturbation Linearization (SPL) focuses on the dynamics close to an equilibrium point to generate corresponding inputs valid around that position. State Feedback Linearization (SFL) is applied to the entire region since $\bm{\eta}$ changes with the real-time measurement of states. Partial Feedback Linearization (PFL) is applies part of the state feedback to achieve better performances with reduced noise.

\subsection{High-level Pendulum Control}\label{III_A}
According to Eq. (\ref{pend_theta12_sim}), we set the pendulum state to be $\bm{x_p}\triangleq[\theta_1\ \theta_2\ \dot{\theta}_1\ \dot{\theta}_2]^\mathrm{T}$ and the lower-dimensional high-level input to be $\bm{u}\triangleq[T_Y\ T_Z]^\mathrm{T}\in\mathbb{R}^2$ with respect to Frame $\{P\}$ in Eq. (\ref{state_space}):
\begin{equation*}
\begin{array}{c}
\bm{F(x_p)} =
\begin{bmatrix}
\dot{\theta}_1 & \dot{\theta}_2 & -\frac{(m_cL_g+m_pL_p)gc\theta_1}{J_{py}+m_cL_g^2} & \dot{\theta}_1\dot{\theta}_2tan\theta_1
\end{bmatrix}^\mathrm{T} \\
\bm{G(x_p)} =
\begin{bmatrix}
0 & 0 & \frac{-1}{J_{py}+m_cL_g^2} & 0 \\ 0 & 0 & 0 & \frac{1/c\theta_1}{J_{pz}+m_cL_g^2} 
\end{bmatrix}^\mathrm{T}
\end{array}    
\end{equation*}

\noindent where $\bm{F(x_p)}$ and $\bm{G(x_p)}$ are nonlinear and can be further converted to $\bm{A}$ and $\bm{B}$ in Eq. (\ref{pseudo_linear_form}).

Furthermore, LQI algorithm requires an internal state $\bm{x_m}$ representing the integral of the attitude error. The error state is defined by $\bm{x_e}\triangleq[e_1\ e_2\ \dot{e}_1\ \dot{e}_2]^\mathrm{T}$ where $e_i\triangleq\theta_i-\theta_{id}$ ($i=1,2$) is the difference between current angles and desired angles.
\begin{equation}
    \bm{\dot{x}_m}=[e_1\ e_2]^\mathrm{T}=
    \begin{bmatrix}
        1 & 0 & 0 & 0 \\
        0 & 1 & 0 & 0
    \end{bmatrix}\bm{x_e}\triangleq \bm{C}\bm{x_e}
    \label{augmented_state}
\end{equation}

Therefore, Eq. (\ref{pseudo_linear_form}) can be further augmented with the internal model. The augmented state $\bm{x_{aug}}\triangleq [\bm{x_e}\ \bm{x_m}]^\mathrm{T}$ is used to obtain the optimal control gain matrix.
\begin{equation}
\bm{\dot{x}_{aug}=A_{aug}x_{aug}+B_{aug}\eta}
\label{state_space_augmented_spl}
\end{equation}

\noindent
where $$
\bm{A_{aug}}=
\begin{bmatrix}
\bm{A} & \bm{0} \\
\bm{C} & \bm{0}
\end{bmatrix},
\bm{B_{aug}}=
\begin{bmatrix}
\bm{B} \\ \bm{0}
\end{bmatrix}
$$

In addition, the cost function of the augmented system is defined as:
\begin{equation}
J(\bm{x_{aug}},\bm{\eta})=\int_0^\infty(\bm{x_{aug}}^\mathrm{T}\bm{Q}\bm{x_{aug}}+\bm{\eta}^\mathrm{T}\bm{R}\bm{\eta})dt
\label{aug_LQI_cost}
\end{equation}

\noindent
where $\bm{Q}\in\mathbb{R}^{6\times6}$ and $\bm{R}\in\mathbb{R}^{2\times2}$ are positive definite matrices showing the weights between the system performance and control efforts. Therefore, the optimal controller gain $\bm{K}\in\mathbb{R}^{2\times6}$ is computed by solving the Riccati Equation with weighed $\bm{Q}$ and $\bm{R}$. The solution is a self-adjoint matrix $\bm{P}\in\mathbb{R}^{6\times6}$ and the actual input $\bm{u}$ can be obtained by the following equation and its relationship with $\bm{\eta}$.
\begin{equation}
\left.
\begin{array}{l}
    \bm{K} = \bm{R}^{-1}\bm{B_{aug}}^\mathrm{T}\bm{P} \\
    \bm{\eta} = -\bm{K}\bm{x_{aug}}
\end{array}
\right.
\label{u_state}
\end{equation}

Since the behavior of different control channels' intersection is not necessarily restricted in our case, $\bm{Q}$ and $\bm{R}$ are set to be diagonal for simplicity. Although the elements in matrices $\bm{A}$ and $\bm{B}$ vary in different methods, they are time-invariant for LQI controller design in each case.

\subsubsection{Small Perturbation Linearization (SPL)}\label{III_A1}

To simplify the model for LQI method, the system can be linearized at some equilibrium point where $\bm{\dot{x}_{eq}}=\bm{0}$. We compute Taylor Expansion for both sides and omit higher order terms in Eq. (\ref{state_space}):
\begin{equation}
\bm{\dot{x}_e} = (\frac{\partial\bm{F}}{\partial\bm{x}}\Big|_{\bm{x_{eq}}}+\frac{\partial\bm{G}}{\partial\bm{x}}\Big|_{\bm{x_{eq}}}\bm{u_{eq}})\bm{x_e} + \bm{G(x_{eq})}\bm{u_e}
\label{base_dynamics_spl}
\end{equation}

\noindent
where $\bm{x_e}\triangleq\bm{x}-\bm{x_{eq}}$ and $\bm{u_e}\triangleq\bm{u}-\bm{u_{eq}}$ stand for the state error and the input error, respectively. $\bm{x_{eq}}$ is determined by regulation specifications and $\bm{u_{eq}}$ is obtained by the equilibrium dynamics:
\begin{equation}
\bm{0} = \bm{F(x_{eq})} + \bm{G(x_{eq})}\bm{u_{eq}}
\label{equi_dynamics_spl}
\end{equation}

In consequence, Eq. (\ref{base_dynamics_spl}) can be converted to a linear form corresponding to Eq. (\ref{pseudo_linear_form}):
\begin{gather*}
\bm{A}=
\begin{bmatrix}
    0 & 0 & 1 & 0 \\
    0 & 0 & 0 & 1 \\
    \frac{(m_cL_g+m_pL_p)gs\theta_{1\_eq}}{J_{py}+m_cL_g^2} & 0 & 0 & 0\\
    0 & 0 & 0 & 0
\end{bmatrix}, \\
\bm{B}=
\begin{bmatrix}
0 & 0 \\  0 & 0 \\ \frac{-1}{J_{py}+m_cL_g^2} & 0 \\ 0 & \frac{1/c\theta_{1\_eq}}{J_{pz}+m_cL_g^2} 
\end{bmatrix},
\bm{\eta}=\bm{u_e}
\end{gather*}
\noindent
where the actual input $\bm{u} = \bm{\eta} + \bm{u_{eq}}$.


\subsubsection{State Feedback Linearization (SFL)}\label{III_A2}
To make Eqs (\ref{state_space}) as a linear system in Eq. (\ref{pseudo_linear_form}) equal, it is partitioned:
\begin{equation}
\begin{array}{l}
\bm{\dot{x}_p} = \bm{F(x_p)}+\bm{G(x_p)u} = 
\begin{bmatrix}
    \bm{F_1x_p} \\ \bm{F_2(x_p)}
\end{bmatrix} + 
\begin{bmatrix}
    \bm{G_1u} \\ \bm{G_2(x_p)u}
\end{bmatrix}
\\ 
\end{array}
\label{pseudo_dynamics_sfl}
\end{equation}

\noindent
where $\bm{F_1}$ and $\bm{G_1}$ represent the linear part while $\bm{F_2(x_p)}$ and $\bm{G_2(x_p)}$ contain nonlinear terms:
\begin{equation*}
\bm{F_1} = 
\begin{bmatrix}
    0 & 0 & 1 & 0 \\ 0 & 0 & 0 & 1
\end{bmatrix},\ 
\bm{G_1} = 
\begin{bmatrix}
    0 & 0 \\ 0 & 0
\end{bmatrix}
\end{equation*}

Therefore, an equivalent system can be obtained:
\begin{equation}
\bm{\dot{x}_p} = 
\begin{bmatrix}
    \bm{F_1} \\ \bm{A_2}
\end{bmatrix}\bm{x_p} + 
\begin{bmatrix}
    \bm{G_1} \\ \bm{B_2}
\end{bmatrix}\bm{\eta}
\end{equation}

\noindent
where 
\begin{equation}
    \bm{u} = \bm{G_2^{-1}(x_p)}[\bm{A_2x_p} + \bm{B_2\eta} - \bm{F_2(x_p)}]    
\end{equation}

\subsubsection{Partial Feedback Linearization (PFL)}\label{III_A3}
The velocity feedback in SFL is not available by measurement.  A common method is to employ a high sampling rate differentiator of the the position measurements. However, the position encoder quantization may introduce significant noises degrading performance or stability.  Partial state feedback may be tailored to replace selective velocity feedback terms by the reference command values.  In the present case, the nonlinear gyroscopic term in Eq. (\ref{pseudo_dynamics_sfl}) is substituted, rendering:
$$
\bm{F_2(x_p)} = 
\begin{bmatrix}
\frac{-(m_cL_g+m_pL_p)gc\theta_1}{J_{py}+m_cL_g^2} &
\dot{\theta}_{1d}\dot{\theta}_{2d}tan\theta_1
\end{bmatrix}^\mathrm{T}
$$
\noindent
where the actual input $\bm{u}$ is processed similar to SFL method. $\dot{\theta}_{1d}$ and $\dot{\theta}_{2d}$ are now the commanded values as the feedforward compensation instead of the feedback measurements.

\subsection{Low-level Gimbal Actuator Control}\label{III_B}

As shown in Eq. (\ref{gimbal_alpbet}), the gimbal dynamics can be rearranged in state-space like Eq. (\ref{state_space}). $\bm{x_g}\triangleq[\alpha\ \beta\ \dot{\alpha}\ \dot{\beta}]^\mathrm{T}$ is a newly defined state vector and $\bm{u}\triangleq[\tau_\alpha\ \tau_\beta]^\mathrm{T}\in\mathbb{R}^2$ is with respect to Frame $\{G\}$: 
\begin{equation*}
\begin{array}{c}
\bm{F(x_g)} =
\begin{bmatrix}
\dot{\alpha} & \dot{\beta} & \frac{[(J_{cx}+J_{cy}-J_{cz})s\beta-J_{cz}c\beta]\dot{\alpha}\dot{\beta}}{J_{cx}c\beta+J_{cz}s\beta} & \frac{(J_{cz}-J_{cx})s\beta c\beta\dot{\alpha}^2}{J_{cy}}
\end{bmatrix}^\mathrm{T} \\
\bm{G(x_g)} =
\begin{bmatrix}
0 & 0 & \frac{c\beta+s\beta}{J_{cx}c\beta+J_{cz}s\beta} & 0 \\ 0 & 0 & 0 & \frac{1}{J_{cy}} 
\end{bmatrix}^\mathrm{T}
\end{array}    
\end{equation*}

Although we could employ similar control design as discussed in Section \ref{III_A}, a cascaded PID controller, which is commanly applied to electric motor control,  is preferred with concrete and physical meanings for tuning the control gains. The relationship between the dynamic characteristics and the PID gains are more comprehensible to students. 
\begin{figure*}[htbp]
\centerline{\includegraphics[scale=0.4]{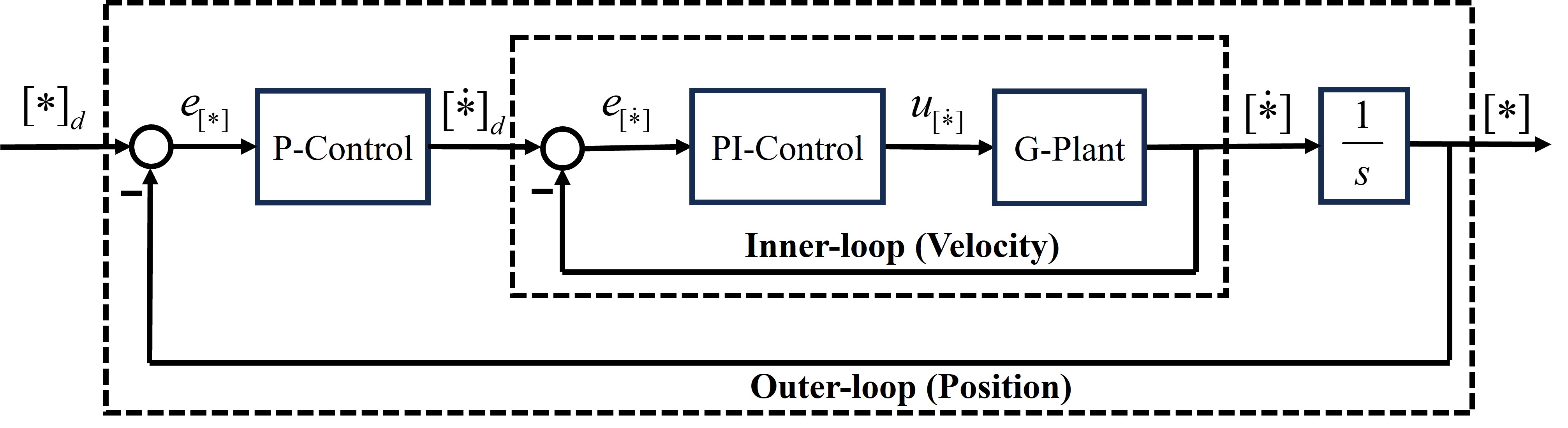}}
\caption{Block diagram of the 2-DoF gimbal. The outer loop tracks the position with a lower bandwidth while the inner loop tracks the velocity with a higher bandwidth for stability. $(*=\alpha,\beta)$}
\label{Low_level_diagram}
\end{figure*}

Emulating second-order linear dynamics, the equivalent moments of inertia defined in $\alpha$ and $\beta$ directions, $J_\alpha$ and $J_\beta$, are cascaded with PI velocity and P position loop error feedback look (Fig. \ref{Low_level_diagram}):
\begin{equation}
\left.
\begin{array}{l}
    \dot{\alpha}_d = k_{P_\alpha}e_\alpha \rightarrow \Ddot{\alpha}_d = k_{P_{\dot{\alpha}}}e_{\dot{\alpha}}+k_{I_{\dot{\alpha}}}\displaystyle\int e_{\dot{\alpha}}dt \\
    \dot{\beta}_d = k_{P_\beta}e_\beta \rightarrow \Ddot{\beta}_d = k_{P_{\dot{\beta}}}e_{\dot{\beta}}+k_{I_{\dot{\beta}}}\displaystyle\int e_{\dot{\beta}}dt
\end{array}
\right.
\label{alpha_beta_PID}
\end{equation}

\noindent
where $k_{P_{[*]}}$, $k_{P_{[\dot{*}]}}$ and $k_{I_{[\dot{*}]}}$ are constant gains. $e_{[*]}\triangleq[*]_d-[*]$ and $e_{[\dot{*}]}\triangleq[\dot{*}]_d-[\dot{*}]$ are corresponding to errors in the position and the angular velocity, respectively. $(*=\alpha,\beta)$

\subsubsection{Small Perturbation Linearization (SPL)}\label{III_B1}
Similar to Section \ref{III_A1}, the error dynamics can be linearized at some equilibrium point where $\dot{\alpha}=\dot{\beta}=0$. Substituting them and $\beta_{eq}$ into Eq. (\ref{state_space}) yields:
\begin{equation}
\left.
\begin{array}{l}
    \Ddot{\alpha}_d = \frac{c\beta_{eq}+s\beta_{eq}}{J_{cx}c\beta_{eq}+J_{cz}s\beta_{eq}}\tau_\alpha \\
    \Ddot{\beta}_d = \frac{1}{J_{cy}}\tau_\beta
\end{array}
\right.
\rightarrow
\left.
\begin{array}{l}
    J_\alpha= \frac{J_{cx}c\beta_{eq}+J_{cz}s\beta_{eq}}{c\beta_{eq}+s\beta_{eq}} \\
    J_\beta=J_{cy}
\end{array}
\right.
\label{ELA_LL}
\end{equation}

\noindent
where $\beta_{eq}$ is the equilibrium position and it is a standard double integrator system regardless of $\alpha_{eq}$.

$J_\beta$ is a constant based on the inertia of the quadcopter while $J_\alpha$ is determined by where $\beta_{eq}$ is. It only works well within the region around the equilibrium point. The actual input $\bm{u} = [ J_\alpha\Ddot{\alpha}_d\ J_\beta\Ddot{\beta}_d]^\mathrm{T}$ where $\Ddot{\alpha}_d$ and $\Ddot{\beta}_d$ come from error dynamics in Eq. (\ref{ELA_LL}).

\subsubsection{State Feedback Linearization (SFL)}\label{III_B2}
Similar to Section \ref{III_A2}, the state feedback linearization yields a double integrator plant with the desired acceleration as the linear plant input $\bm\eta$:
\begin{equation}
\begin{array}{l}
    \Ddot{\alpha}_d = \frac{[(J_{cx}+J_{cy}-J_{cz})s\beta-J_{cz}c\beta]\dot{\alpha}\dot{\beta}}{J_{cx}c\beta+J_{cz}s\beta} + \frac{c\beta+s\beta}{J_{cx}c\beta+J_{cz}s\beta}\tau_\alpha \\
    \Ddot{\beta}_d = \frac{(J_{cz}-J_{cx})s\beta c\beta\dot{\alpha}^2}{J_{cy}} + \frac{1}{J_{cy}}\tau_\beta
\end{array}
\label{SFL_LL}
\end{equation}

\noindent
where $\dot{\alpha}$ and $\dot{\beta}$ are state variables. The system does not neglect the gyroscopic effect, therefore, allowing for accurate compensation of the nonlinear terms in the feedback control. The actual input is defined accordingly:
\begin{equation}
\begin{array}{l}
    J_\alpha(\bm{x_g})= \frac{J_{cx}c\beta+J_{cz}s\beta}{c\beta+s\beta} \\
    J_\beta=J_{cy}
\end{array}
\rightarrow
\begin{array}{l}
    \tau_\alpha = J_\alpha\Ddot{\alpha_d} + \gamma(\bm{x_g})\\
    \tau_\beta = J_\beta\Ddot{\beta_d} + \phi(\bm{x_g})
\end{array}
\label{control_input_sfl}
\end{equation}

\noindent
where $\gamma(\bm{x_g})$ and $\phi(\bm{x_g})$ absorbs coupling effects. Different from the analysis in Section \ref{III_B1}, $J_\alpha$ becomes a function of $\beta$ in this case while $J_\beta$ still remains constant. 

\subsubsection{Partial Feedback Linearization (PFL)}\label{III_B3}
To alleviate the degradation by noisy velocity feedback measurement, which is from IMU data with internal estimation algorithms, PFL replaces the velocity measurements by the commanded velocity in computing the nonlinear compensation terms $\gamma(\bm{x_g})$ and $\phi(\bm{x_g})$.
It should be noted that the linear velocity feedback terms are still intact for the closed-loop error dynamics.
\begin{equation}
\begin{array}{l}
    \Ddot{\alpha}_d = \frac{[(J_{cx}+J_{cy}-J_{cz})s\beta-J_{cz}c\beta]\dot{\alpha}_d\dot{\beta}_d}{J_{cx}c\beta+J_{cz}s\beta} + \frac{c\beta+s\beta}{J_{cx}c\beta+J_{cz}s\beta}\tau_\alpha \\
    \Ddot{\beta}_d = \frac{(J_{cz}-J_{cx})s\beta c\beta\dot{\alpha}_d^2}{J_{cy}} + \frac{1}{J_{cy}}\tau_\beta
\end{array}
\label{PFL_LL}
\end{equation}

\noindent
where the actual input $\bm{u}$ is processed similar to SFL method. $\dot{\alpha}_d$ and $\dot{\beta}_d$ are now the commanded value as the feedforward compensation instead of the feedback state.

\subsection{Control Allocation and Mapping for Over-Actuation}\label{III_C}

This 4-DoF actuation can be set up to control the 2-DoF pendulum angles. We consider a hierarchical control structure, which is common in many mechanical systems. For example, a robot manipulator has a high-level motion control that gives joint velocity commands for a low-level joint motor control to follow by a servo control. In terms of the proposed platform, the high-level algorithm controls the pendulum angles by giving torque commands to the low-level actuator. The low-level actuator responses to such commands by controlling the force and attitude of the quadcopter with $f$, $\alpha$ and $\beta$.

According to Section \ref{III_A}, $\prescript{P}{}{\bm{u}}$ is obtained from $\bm{\eta}$ in Eq. (\ref{u_state}) but we need to convert it to $\prescript{G}{}{\bm{y}}$ in Section \ref{III_B} for the actuator. It is achieved by allocating an intermediate thrust vector $\prescript{G}{}{\bm{F}}$ appropriately.

Since $F_y$ and $F_z$ are given by Eq. (\ref{pend_torque}), $F_x$ is redundant due to the over-actuation. There are many solutions to $F_x$ but the stability and safety when hovering must be the priority. In this paper, it is determined by where $\theta_1$ is: When $\theta_1$ is away from $\pi/2$, the desired $\beta_d$ is set to $\theta_1$ to keep the quadcopter horizontal all the time. Combining Eqs. (\ref{F_G}) and (\ref{f_alp_bet}) yields:
\begin{equation}
    F_x\triangleq |tan\theta_1|\sqrt{F_y^2+F_z^2}
    \label{Fx_normal}
\end{equation}

Otherwise, the actuator falls into the gimbal lock singularity near $\theta_1=\pi/2$. Assuming that the gimbal moves slowly with dynamic equilibrium, the net force along +Z axis in Frame $\{W\}$ should be zero on condition that the reaction and support force at the base is neglected.
\begin{equation}
    F_xs\theta_1+F_zc\theta_1=(m_c+m_p)g
    \label{Fx_singular}
\end{equation}

\subsection{Input and Output Constraints}\label{III_D}

In the high-level controller, $\theta_1$ and $\theta_2$ are limited in a hemispherical workspace where $\theta_1\in[0,\pi]$ and $\theta_2\in[0,2\pi]$. When $\theta_1$ reaches $\pi/2$, it is defined as the singularity point where the system loses controllability according to the state space matrices in Section \ref{III_A1}. 

The auxiliary matrix does not have full rank and the null-space rows are corresponding to $\theta_2$ and $\dot{\theta}_2$. Therefore, $\theta_2$ becomes uncontrollable and $\theta_1$ is aligned with $\beta$. $F_x$ and $F_z$ are not affected, leading $\theta_1$ to remain controllable in spite of the singularity.

To solve the singularity case, within the range of $\theta_1\in\pi/2\pm\delta$ where we define $\delta \triangleq \pi/9$, $F_y$ is manually set to zero to avoid infinite input calculation. $T_Z$ can be any value but the control channel is turned off so that it can keep stable by friction.

\section{Gimbal Actuator Control}\label{IV}

To verify the controller design in Section \ref{III_B}, a Simscape-Multibody plant is built and configured in MATLAB Simulink to visualize the gimbal dynamics under feedback control. The simulation model includes every rigid body of the imported CAD objects and assembly, which does not assume that the actuator is a point mass as in Eq. (\ref{gimbal_Np}). Since the angle axis $\alpha$ has a significantly larger inertia than the angle $\beta$, the PID adjustment is only performed on $\alpha$ but applied to both axes. The ceiling of the generated thrust is modeled, which is useful in tuning the gains to compromise between closed-loop performance and actuator saturation (Table \ref{copter_param}). 
\begin{table}[width=\linewidth,pos=h]
\caption{Model Parameters and Physical Variables of the Copter}\label{copter_param}
\begin{tabular*}{\tblwidth}{C|C|C|C}
\hline
Symbols & Descriptions & Values & Units\\
\hline
$m_c$ & Copter mass (with battery) & 26 & g \\
\hline
$g$ & Gravity & 9.81 & $m\cdot s^{-2}$ \\
\hline
$f$ & Thrust capability & 0.6 & N \\
\hline
$J_{cx},J_{cy}$ & Moment of inertia (xy-axis) & 166 & $g\cdot cm^2$ \\
\hline
$J_{cz}$ & Moment of inertia (z-axis) & 293 & $g\cdot cm^2$ \\
\hline
\end{tabular*}
\end{table}

However, some properties are hard to model and have to be simplified in the simulation. The friction due to the gimbal rotation is neglected and the aerodynamics of each propeller is assumed to only generate a thrust force and a resistance torque. The aforementioned continuous-time controllers are realized in the discrete-time with the same sampling rate as the experiment. The desired command for the gimbal actuator is specified and sent wirelessly to the quadcopter's rotor controller by a radio (2.4 GHz Crazyradio PA). It receives the reference commands and closes the corresponding servo-loop at a rate of 500 Hz.

\subsection{Step and Ramp Response}

\begin{figure}[htbp]
    \centering
    \begin{subfigure}[b]{0.35\textwidth}
        \centering
        \includegraphics[width=\textwidth]{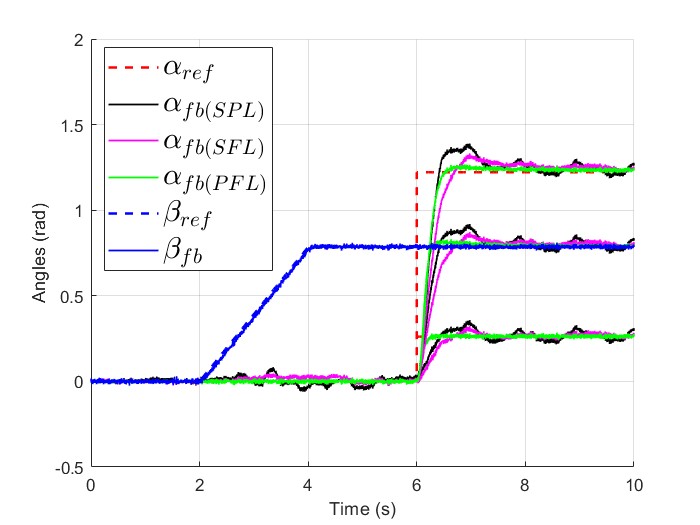}
        \caption{Simulation}
    \end{subfigure}
    \begin{subfigure}[b]{0.35\textwidth}
        \centering
        \includegraphics[width=\textwidth]{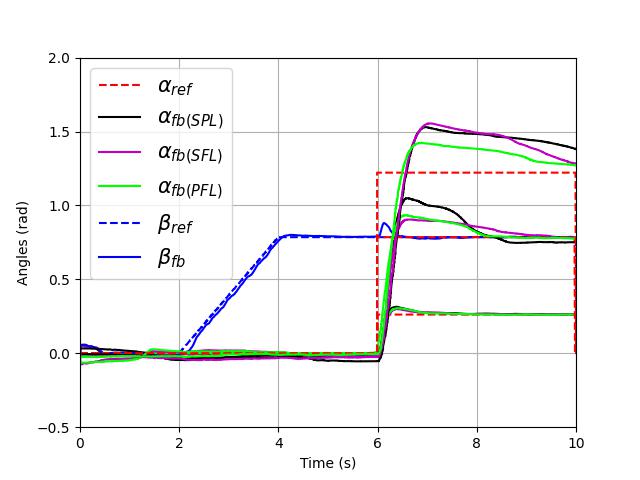}
        \caption{Experiment}
    \end{subfigure}
    \caption{Closed-loop performance of the gimbal. The ramp response of $\beta$ ($0^\circ\rightarrow45^\circ$) is followed by the step response of $\alpha$ ($15^\circ/45^\circ/70^\circ$) with different magnitudes. There exists a small perturbation of $\beta$ at $t=6$ when $\alpha$ signal is triggered due to the coupling effect from the model dynamics.}
    \label{low_level_from45_perf}
\end{figure}

Considering that the equilibrium position of the gimbal actuator is $\alpha = \beta = 0$ at the initial state, SPL controller is linearized at $\beta_e=0$ while PFL controller regards $J_\alpha$ as a function of $\beta$. Furthermore, the coupling via velocities is taken into account by SFL controller but not included in PFL controller (Fig. \ref{low_level_from45_perf}).

The difference between them is negligible if the step is small ($15^\circ$) but becomes noticeable when the step is larger ($45^\circ$ and $70^\circ$). Compared to the simulation results, $\alpha$ angle's step responses have similar rise time but higher overshoot and longer settling time at large steps. This can be explained by the processing difference between the simulation and the experiment: In the simulation, $\alpha$ and $\beta$ angles are directly recorded by joint blocks sensing. In the experiment, they are estimated from the IMU data created by an internal Kalman Filter, which introduces unmodeled dynamics in the feedback control loop and causes actuator saturation.
\begin{table}[width=\linewidth,pos=h]
\caption{Gimbal Loop Transient Performance ($\alpha_{ref}=45^\circ$)}\label{gimbal_step}
\begin{tabular*}{\tblwidth}{C|C|C|C}
\hline
Simulation & Rise Time (s) & Overshoot (\%) & Settling Time (s)\\
\hline
SPL & 0.30 & 17.1 & 2.12 \\
SFL & 0.41 & 11.0 & 1.24 \\
PFL & 0.16 & 5.7 & 0.63 \\
\hline
Experiment & Rise Time (s) & Overshoot (\%) & Settling Time (s)\\
\hline
SPL & 0.19 & 33.6 & 1.89 \\
SFL & 0.23 & 18.9 & 2.32 \\
PFL & 0.21 & 15.2 & 1.91 \\
\hline
\end{tabular*}
\end{table}

Since SPL does not work well away from the equilibrium position and SFL is negatively affected by the sensor noise, PFL performs best in the simulation tests, as depicted in Table \ref{gimbal_step}. Therefore, PFL is adopted as the low-level controller in the following high-level simulations and experiments.

\subsection{Rotor Inputs Comparison}

\begin{figure}[htbp]
     \centering
     \begin{subfigure}[b]{0.23\textwidth}
         \centering
         \includegraphics[width=\textwidth]{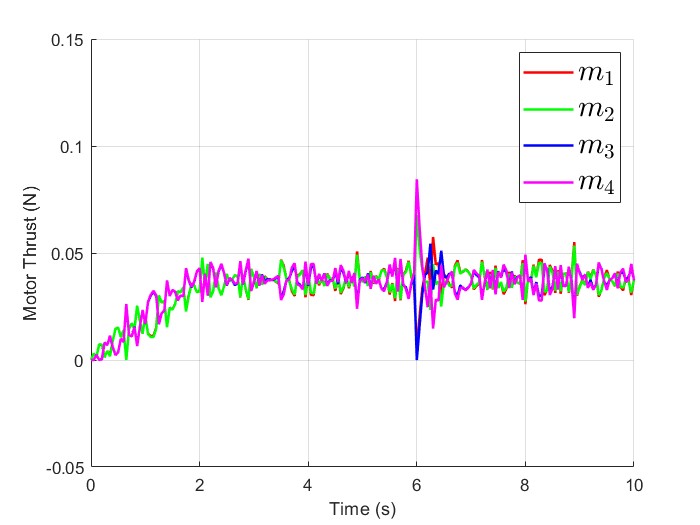}
         \caption{SPL Controller}
         \label{low_level_sim_from45_45_thrust_ELA}
     \end{subfigure}
     \begin{subfigure}[b]{0.23\textwidth}
         \centering
         \includegraphics[width=\textwidth]{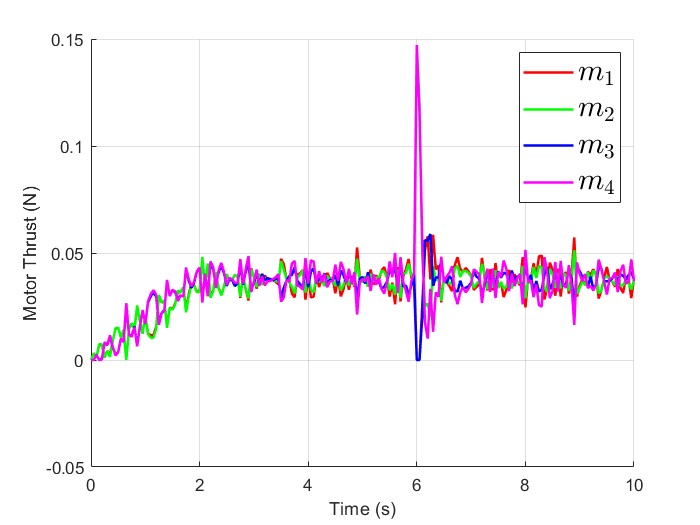}
         \caption{SFL Controller}
         \label{low_level_sim_from45_45_thrust_NSF}
     \end{subfigure}
     \begin{subfigure}[b]{0.23\textwidth}
         \centering
         \includegraphics[width=\textwidth]{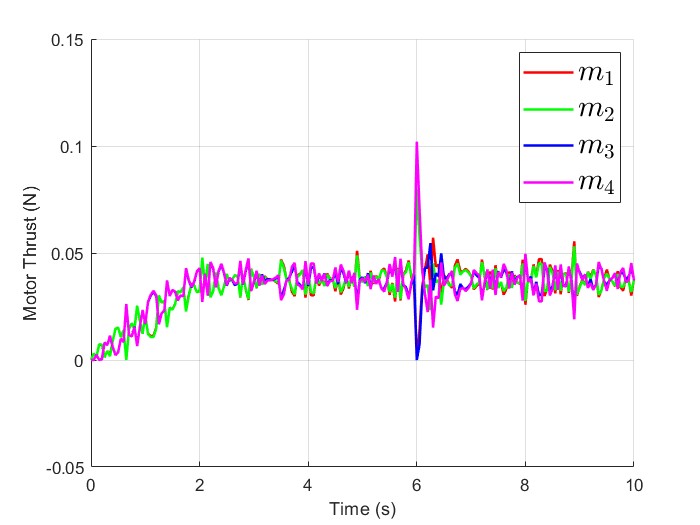}
         \caption{PFL Controller}
         \label{low_level_sim_from45_45_thrust_PSL}
     \end{subfigure}
     \caption{The thrust commands of the rotor corresponding to different controllers when the step signal $\alpha_{ref}=45^\circ$ is applied at $t=6s$. Actuator commands are within the saturation limit of $0.1472N$.}
    \label{low_level_sim_from45_45_thrust}
\end{figure}

The PID controller as in Eq. (\ref{alpha_beta_PID}) is manually tuned in the simulation to optimize and balance the rise time, overshoot, and settling time without triggering the actuator saturation limit. The difference of rotor inputs between three controllers for a 45-degree step change in $\alpha$ is shown (Fig. \ref{low_level_sim_from45_45_thrust}).

\begin{figure}[htbp]
\centerline{\includegraphics[scale=0.4]{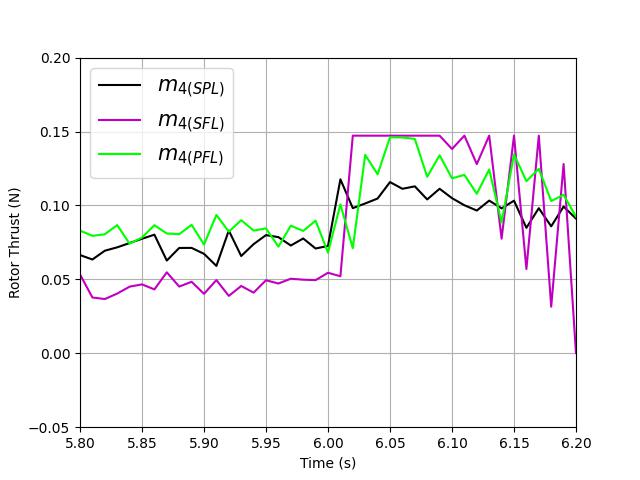}}
\caption{The thrust command of the rotor $m_4$ when $\alpha_{ref}=45^\circ$. They are zoomed in around $t=6$ when the step signal is introduced.}
\label{low_level_exp_from45_45_thrust_ELA_NSF_PSL}
\end{figure}

Based on the simulation results, the gains are implemented in the experiment and one of the motor thrusts is recorded (Fig. \ref{low_level_exp_from45_45_thrust_ELA_NSF_PSL}). Barring from the actuator saturation hampering the settling time, the performance of PFL is superior to SPL and SFL, which is consistent with the simulation result. Since the high-level pendulum control commonly gives small incremental low-level commands, the case of large overshoot and settling time is unlikely to occur. This is supported by the fact that the transient performance for a 15-degree step change agrees closer to that of the simulation (Fig. \ref{low_level_from45_perf}).

\section{Pendulum Control}\label{V}

The platform simulation is to verify the response of the spherical pendulum. The linearized controller is compared with the nonlinear controllers in the regulation and tracking cases presented next. The same trajectories are implemented in the experiment for validation. Furthermore, The sensor noise from the encoders and the onboard estimator is modeled as white noise with normal distribution. The model parameters and dynamic variables are given in Table \ref{pend_param}.
\begin{table}[width=\linewidth,pos=h]
\caption{Model Parameters and Physical Variables of the Pendulum}\label{pend_param}
\begin{tabular*}{\tblwidth}{C|C|C|C}
\hline
Symbols & Descriptions & Values & Units\\
\hline
$L_g$ & Pendulum length & 33.7 & cm \\
\hline
$L_p$ & Pendulum CoM & 28.5 & cm \\
\hline
$m_p$ & Pendulum mass & 11 & g \\
\hline
$J_{py},J_{pz}$ & Moment of inertia (yz-axis)& 8934.7 & $g\cdot cm^2$ \\
\hline
\end{tabular*}
\end{table}

\begin{figure}[htbp]
\centerline{\includegraphics[scale=0.25]{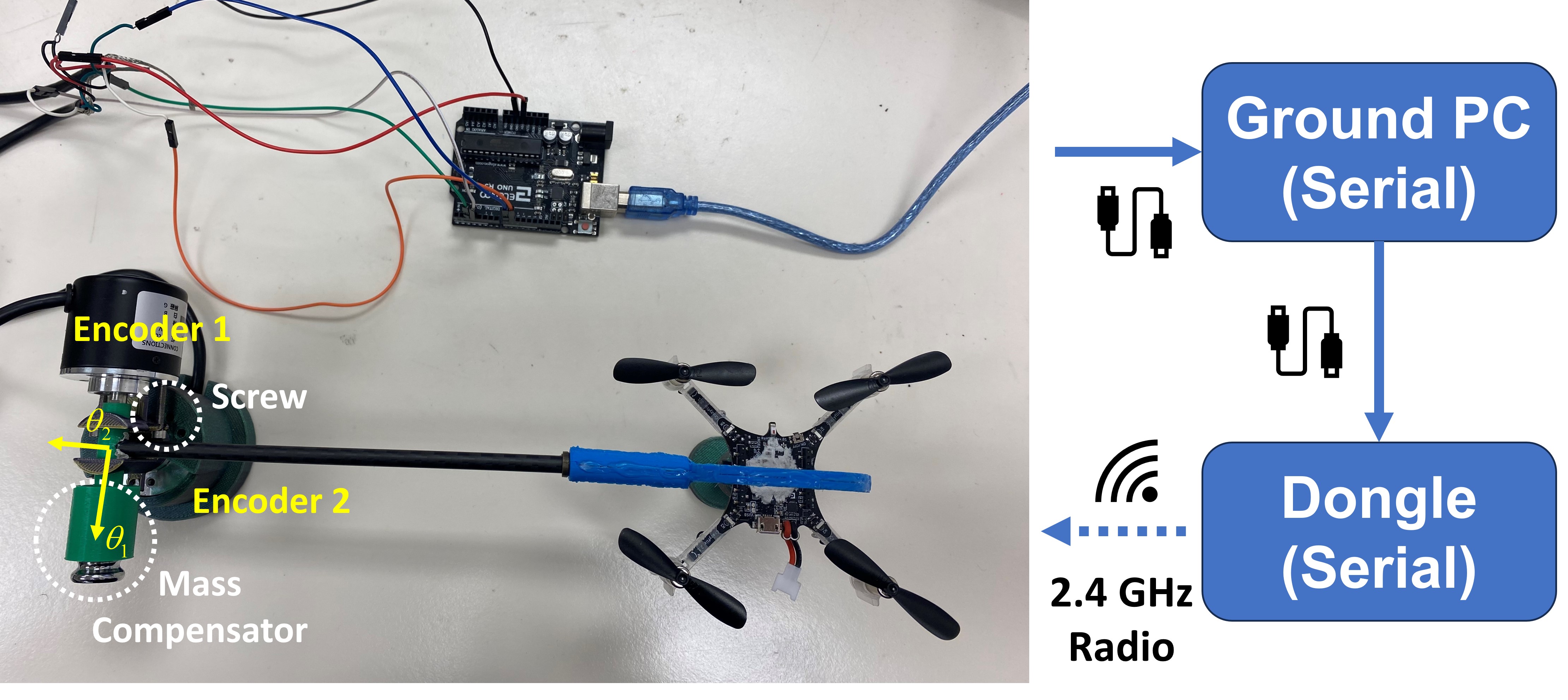}}
\caption{Connection overview and experimental setup with the base and the encoders. $\theta_1$ and $\theta_2$ have free rotation. The high-level commands are sent at 100 Hz while the low-level commands are tracked at 500 Hz}
\label{Connection_Overview}
\end{figure}

In the experimental hardware setup (Fig. \ref{Connection_Overview}), two rotary, incremental encoders are sampled and decoded on an ATmega328 Arduino UNO micro-controller. The encoders provide direct measurements of the horizontal and vertical angles. A single-pole low-pass filter with the cut-off frequency at 20 Hz is cascaded with the encoder feedback to attenuate the signal noise. The high-level controller is implemented on a PC with Intel i7-10750H CPU communicating with the UNO via a serial communication line at 9600 baud rate and 100 Hz. The UNO sends a data packet of two floats (8 bytes) at a rate of 120 Hz.

\subsection{Regulation}\label{V_A}
Since SPL method is only valid within a small region around the equilibrium position, it can be applied to the regulation case. Although SFL and PFL are effective as well, SPL is preferred for the simplified dynamics. In this case, $\theta_{1\_ref}=0$ and $\pi/6$ are corresponding to the equilibrium and non-equilibrium positions, respectively.

\subsubsection{Controller Performance}\label{V_A1}

Since the controller is designed for a locally linearized system at $\theta_{1\_eq}=0$, the regulation at $\pi/6$, though stable for a short time, diverges to infinity in the end. The experimental result is consistent with that of the simulation, indicating that SPL can work pretty well in some cases even in a highly-nonlinear system.
\begin{figure}[htbp]
     \centering
     \begin{subfigure}[b]{0.3\textwidth}
         \centering
         \includegraphics[width=\textwidth]{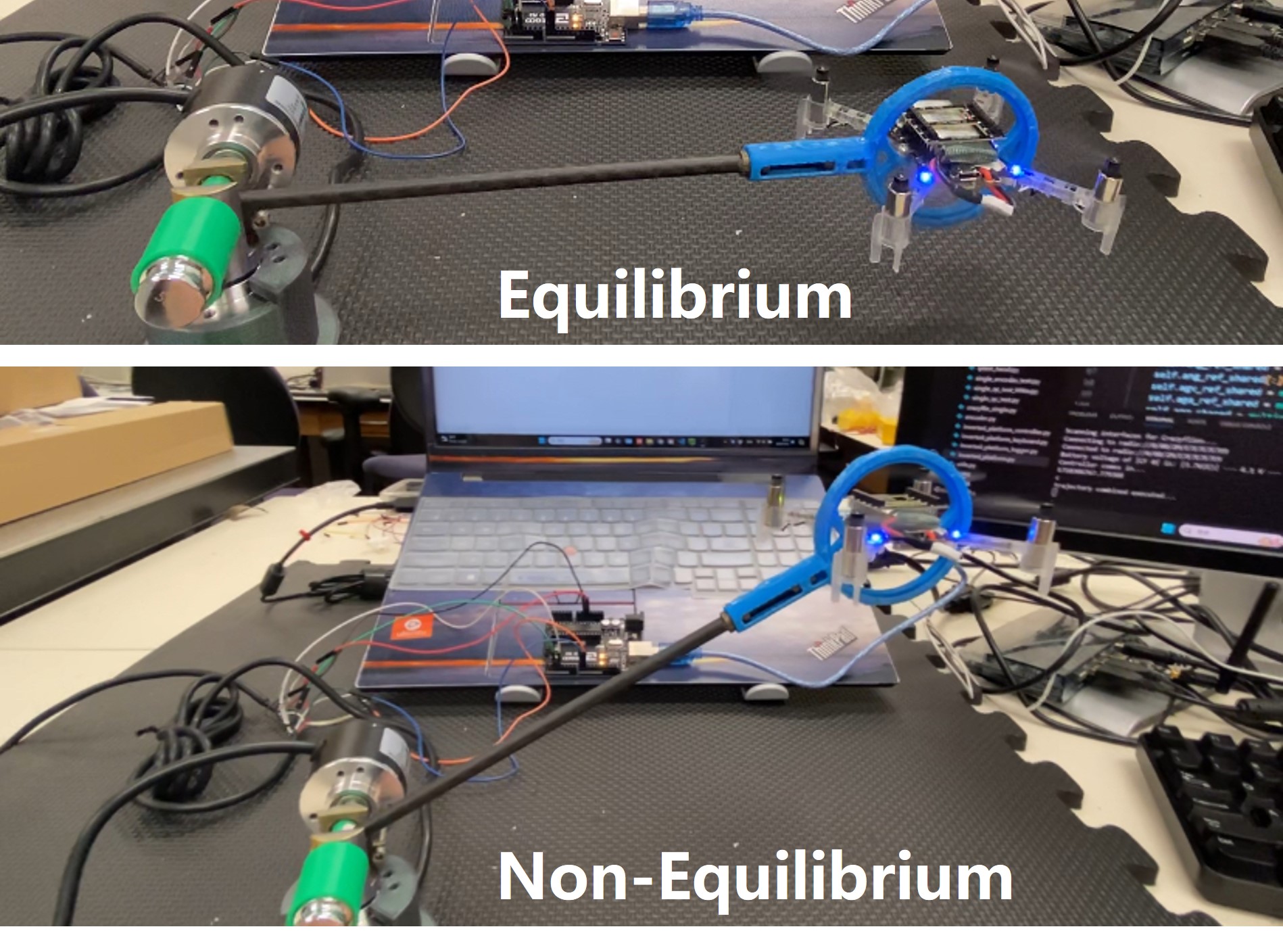}
         \caption{Regulation Trajectory Cases:\\Equilibrium $\theta_{1\_ref}=0$\\Non-Equilibrium $\theta_{1\_ref}=\pi/6$}
     \end{subfigure}
     \begin{subfigure}[b]{0.23\textwidth}
         \centering
         \includegraphics[width=\textwidth]{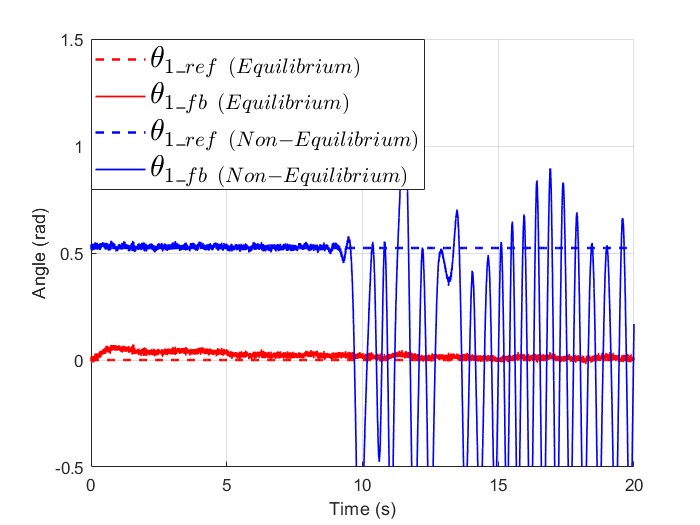}
         \caption{$\theta_1$ Simulation}
     \end{subfigure}
     \begin{subfigure}[b]{0.23\textwidth}
         \centering
         \includegraphics[width=\textwidth]{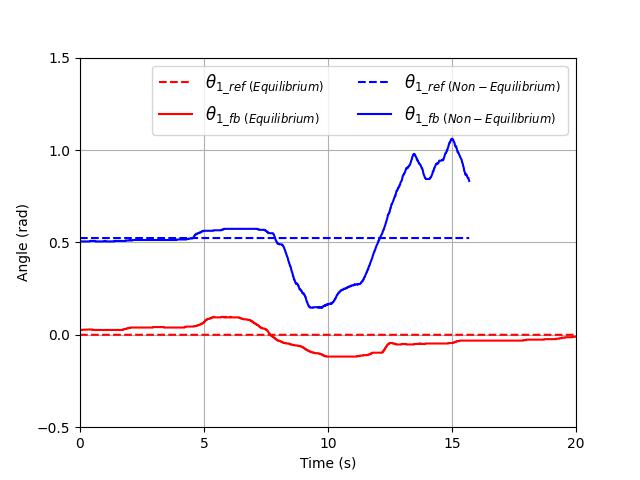}
         \caption{$\theta_1$ Experiment}
     \end{subfigure}
     \begin{subfigure}[b]{0.23\textwidth}
         \centering
         \includegraphics[width=\textwidth]{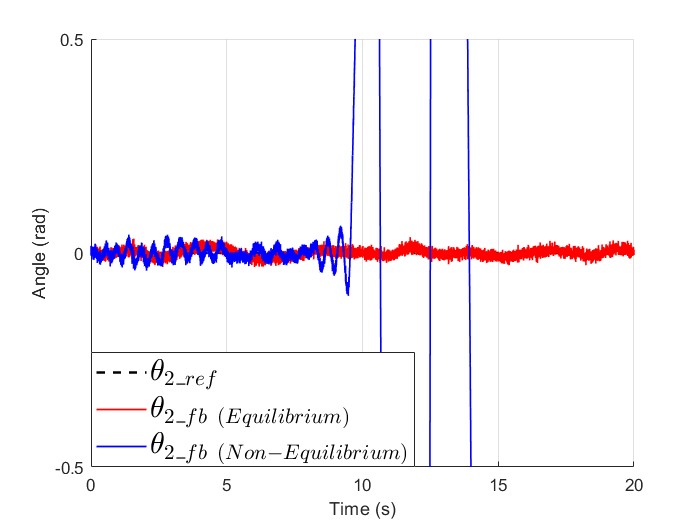}
         \caption{$\theta_2$ Simulation}
     \end{subfigure}
     \begin{subfigure}[b]{0.23\textwidth}
         \centering
         \includegraphics[width=\textwidth]{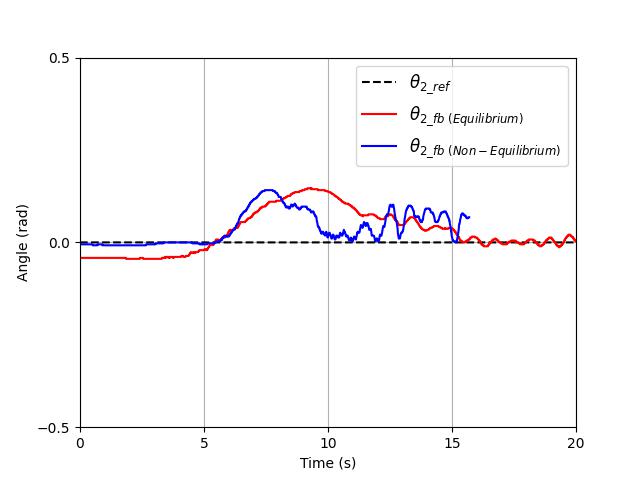}
         \caption{$\theta_2$ Experiment}
     \end{subfigure}
     \caption{$\theta_1$ Regulation performance. $\theta_{1\_ref}$ is set to the equilibrium and non-equilibrium position for 20 seconds while $\theta_{2\_ref}$ remains at $0$ for both cases. The platform is released at $t=5$ and $\theta_1$ can only be maintained in the equilibrium case. The non-equilibrium case fails at around $t=8$ in both the simulation the experiment.}
    \label{high_level_regulation_ELA_perf}
\end{figure}

However, the significant coupling between the azimuth and elevation angle is not evident in the simulation. By contrast, there is a noticeable deviation of $\theta_2$ at $t=6$ in the experiment (Fig. \ref{high_level_regulation_ELA_perf}) due to the fluctuation of $\theta_1$ correspondingly. The RMS errors for all three cases are shown in Table \ref{regu} as a summary.

\begin{table}[width=\linewidth,pos=h]
\caption{RMS Error of Regulation Performance ($\theta_1$ focused)}
\begin{tabular*}{\tblwidth}{|C|C|C|C|C|}
\hline
\multicolumn{2}{|c|}{\diagbox{Case}{Unit (rad)}{Controller}} & \makecell[c]{SPL \\ ($\theta_{1\_eq}=0$)} & SFL & PFL \\
\hline
\multirow{2}{*}{Simulation}& Equilibrium & 0.026 & 0.028 & 0.023 \\
\cline{2-5}
& Non-Equilibrium & Unstable & 0.031 & 0.025 \\
\hline
\multirow{2}{*}{Experiment}& Equilibrium & 0.016 & 0.033 & 0.029 \\
\cline{2-5}
& Non-Equilibrium & Unstable & 0.039 & 0.032 \\
\hline
\end{tabular*}
\label{regu}
\end{table}

\subsubsection{Disturbance Recovery}\label{V_A2}
Apart from verifying the efficacy, the capability of recovering from disturbance is of significance in the controller design. In the stable regulation case at the equilibrium position, we examine the system performance when it undergoes external impulse or step perturbation.

\begin{figure}[htbp]
\centering
    \begin{subfigure}[b]{0.23\textwidth}
         \centering
         \includegraphics[width=\textwidth]{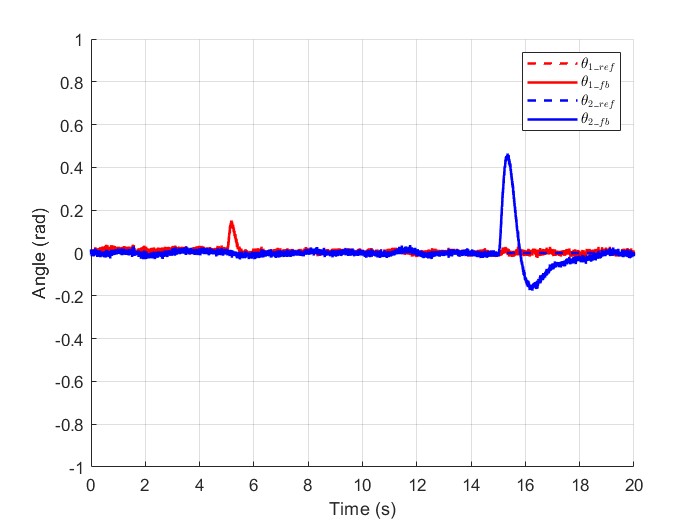}
         \caption{Impulse Disturbance $\tau_d = 0.1N\cdot m$}
         \label{impulse_dist}
    \end{subfigure}
    \begin{subfigure}[b]{0.23\textwidth}
         \centering
         \includegraphics[width=\textwidth]{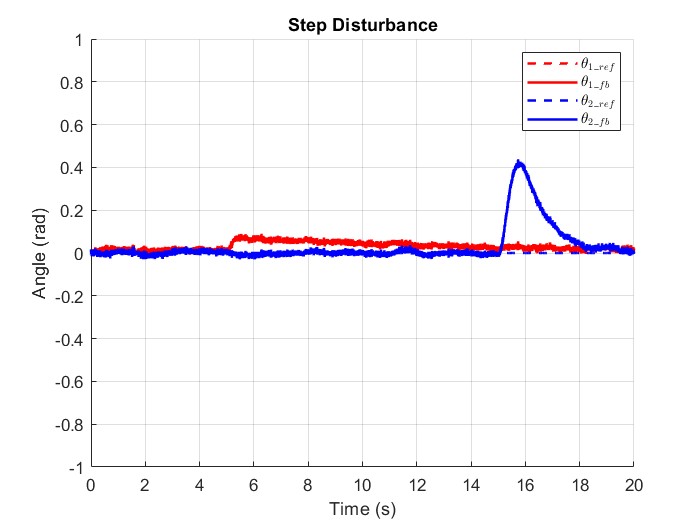}
         \caption{Step Disturbance $\tau_d = 0.02N\cdot m$}
         \label{step_dist}
     \end{subfigure}
\caption{Disturbance Recovery with SPL Controller. The same magnitude of torque is triggered at $t=5$ for $\theta_1$ and $t=15$ for $\theta_2$ during the regulation process between 0 and 20 seconds.}
\label{Disturbance_Recovery}
\end{figure}

Although it is hard to apply disturbance quantitatively in the experiment, it can be solved in the simulation by adding an external torque that is induced at the specified time with precise magnitudes. $\tau_d$ is defined as the disturbance torque applied on the base and it is triggered separately for $\theta_1$ and $\theta_2$ to avoid the coupling effect. Since all the controllers have similar performances in the regulation process, SPL is adopted for simplicity (Fig. \ref{Disturbance_Recovery}).

It is concluded that SPL method handles the impulse disturbance better than the step disturbance with less time to recover. Furthermore, the deviation for $\theta_2$ is notably larger than $\theta_1$, which can be explained by the difference in the recovery mechanism. When the pendulum is regulated around the equilibrium position, the change in $\theta_1$ is achieved by increasing or decreasing the thrust magnitude directly while the change in $\theta_2$ has to be realized by tilting the thrust direction. In general cases, it is more efficient to adjust the thrust magnitude rather than the actuator orientation in the control loop.

\subsection{Tracking below the Inverted Position}\label{V_B}

\begin{figure}[htbp]
\centering
    \begin{subfigure}[b]{0.45\textwidth}
        \centering
        \includegraphics[width=\textwidth]{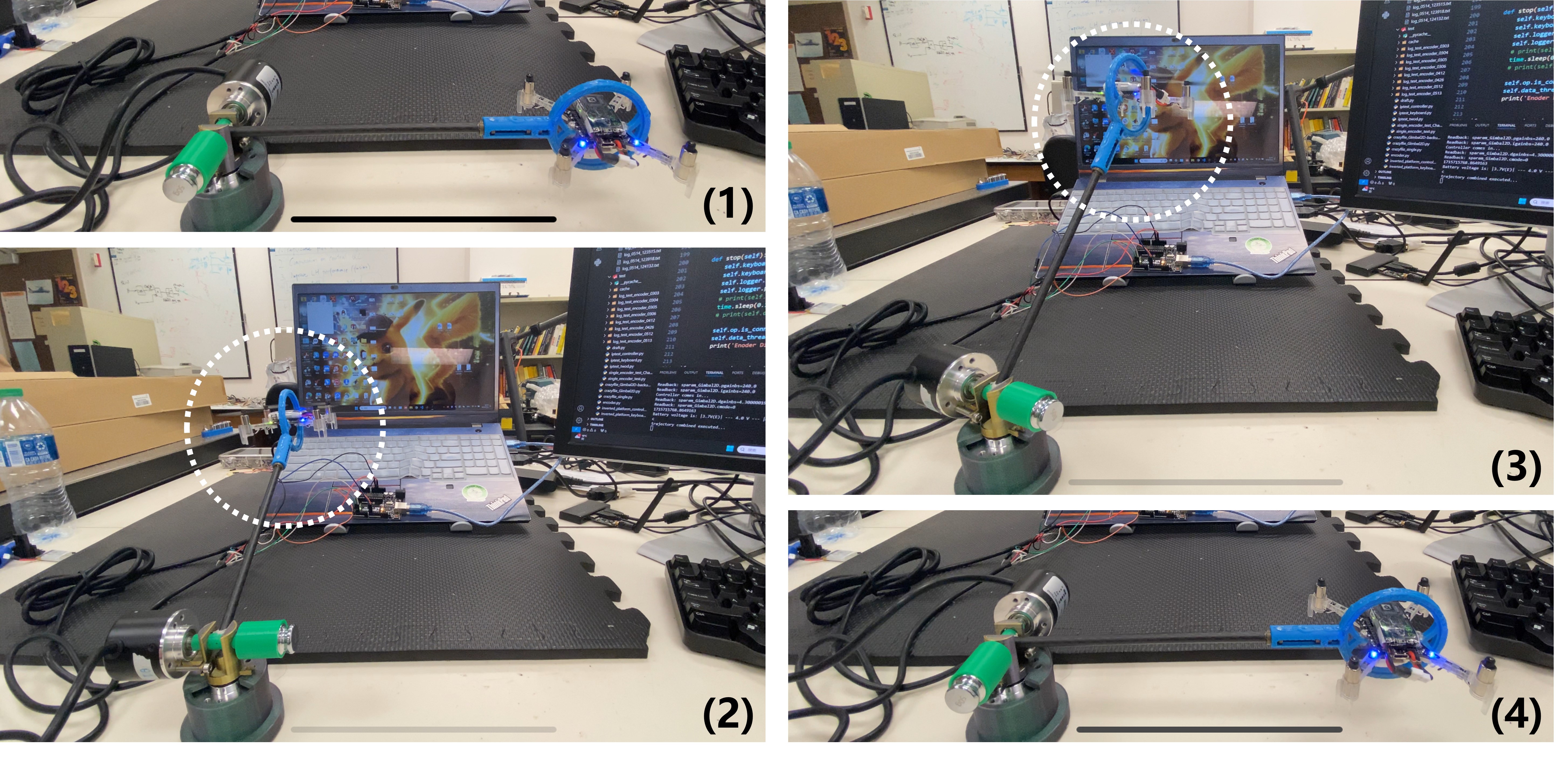}
        \caption{Combined Trajectory Process $(1)\rightarrow(2)\rightarrow(3)\rightarrow(4)$:\\$\theta_{1\_ref}=0\rightarrow\pi/6\rightarrow\pi/4\rightarrow0$\\$\theta_{2\_ref}=0\rightarrow\pi/2\rightarrow\pi/3\rightarrow0$ }
        \label{high_level_exp_combined_ELA_NSF_PSL_setup}
    \end{subfigure}
    \begin{subfigure}[b]{0.23\textwidth}
         \centering
         \includegraphics[width=\textwidth]{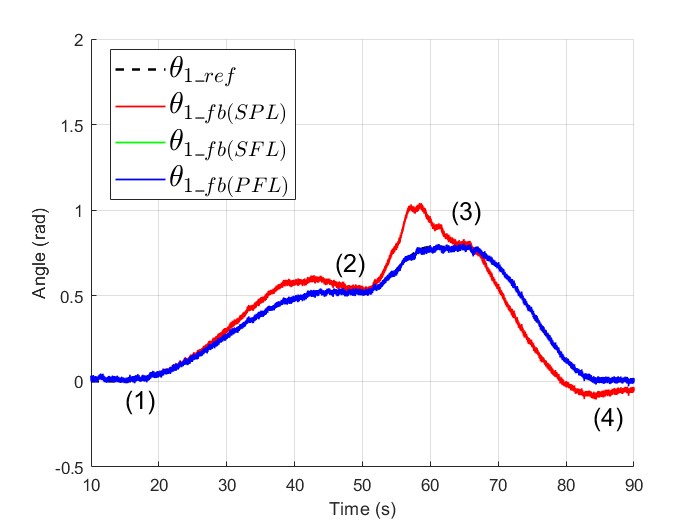}
         \caption{$\theta_1$ Simulation}
         \label{high_level_sim_combined_theta1}
    \end{subfigure}
    \begin{subfigure}[b]{0.23\textwidth}
         \centering
         \includegraphics[width=\textwidth]{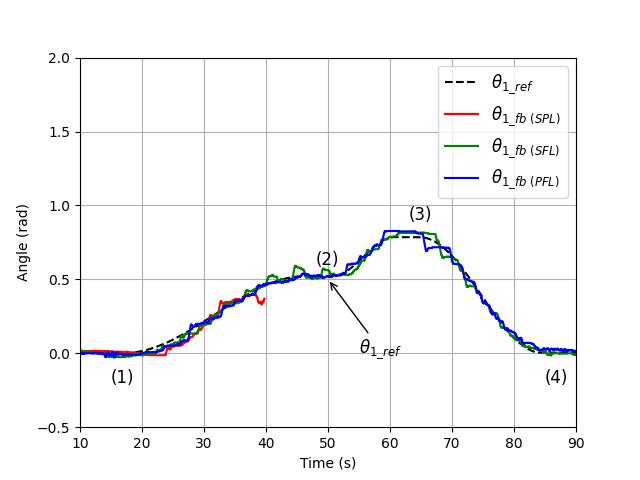}
         \caption{$\theta_1$ Experiment}
         \label{high_level_exp_combined_theta1}
     \end{subfigure}
     \begin{subfigure}[b]{0.23\textwidth}
         \centering
         \includegraphics[width=\textwidth]{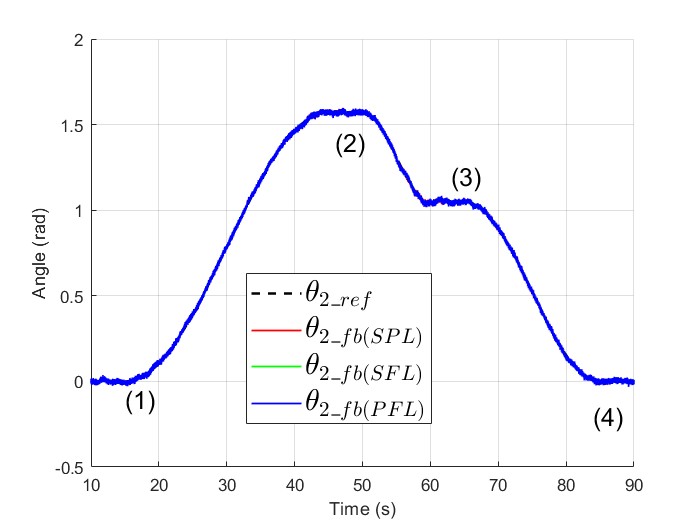}
         \caption{$\theta_2$ Simulation}
         \label{high_level_sim_combined_theta2}
    \end{subfigure}
     \begin{subfigure}[b]{0.23\textwidth}
         \centering
         \includegraphics[width=\textwidth]{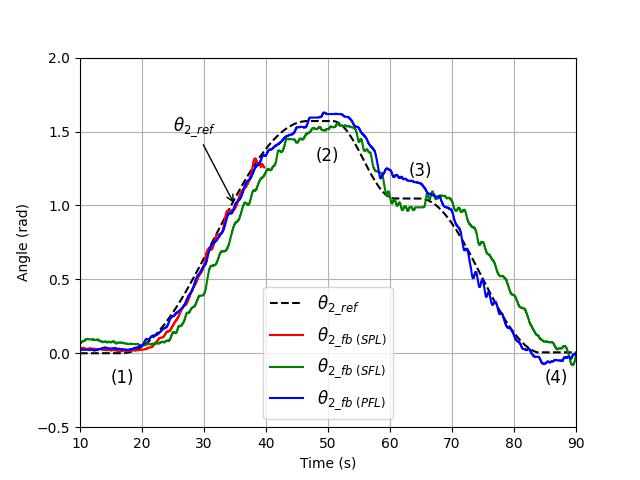}
         \caption{$\theta_2$ Experiment}
         \label{high_level_exp_combined_theta2}
     \end{subfigure}
\caption{Tracking below the Inverted Position: $\theta_1$ and $\theta_2$ are commanded to change simultaneously within the range of the spherical workspace. Two sequences of angles are designed to be tracked between 10 and 90 seconds using different controllers}
\label{Combined Trajectory}
\end{figure}

Tracking of a trajectory without reaching the inverted position is considered in this case. The LQI control weighting matrix per Eq. (\ref{aug_LQI_cost}) are $\bm{Q}=diag(10^4,10^2,0,0,10^2,10^2)$ and $\bm{R} = diag(10^6,10^6)$, which heavily penalizes the running position error and inputs to avoid saturation. To make the transition smoother between angle commands, a sinusoidal trajectory is applied accordingly. All the linear and nonlinear controllers maintain stability and track $\theta_{2\_ref}$ profiles regardless of the coupling effect from the model dynamics. However, $\theta_{1\_ref}$ can only be tracked well with SFL and PFL controllers since SPL is only valid around the equilibrium position at $\theta_{1\_eq}=0$ (Fig. \ref{Combined Trajectory}).

It is demonstrated that the platform can track two angles independently even if there exists a coupling effect between them. SPL controller fails at $t=40$ in the experiment and PFL method outperforms SFL method with less RMS error as shown in Table \ref{track_below_inv}. The thrust command is smoother and the rotor commands are less aggressive away from the singularity region.
\begin{table}[width=\linewidth,pos=h]
\caption{RMS Error of Combined Tracking Performance ($\theta_1$ focused)}
\centering
\begin{tabularx}{\linewidth}{|>{\centering\arraybackslash}X|C|C|C|}
\hline
\diagbox{Case}{Unit (rad)}{Controller} & \makecell[c]{SPL \\ ($\theta_{1\_eq}=0$)} & SFL & PFL \\
\hline
Simulation & 0.107 & 0.017 & 0.014 \\
\hline
Experiment & Unstable & 0.037 & 0.022 \\
\hline
\end{tabularx}
\label{track_below_inv}
\end{table}


\subsection{Tracking towards the Inverted Position}\label{V_C}
\begin{figure}[htbp]
     \centering
     \begin{subfigure}[b]{0.35\textwidth}
     \centering
     \includegraphics[width=\textwidth]{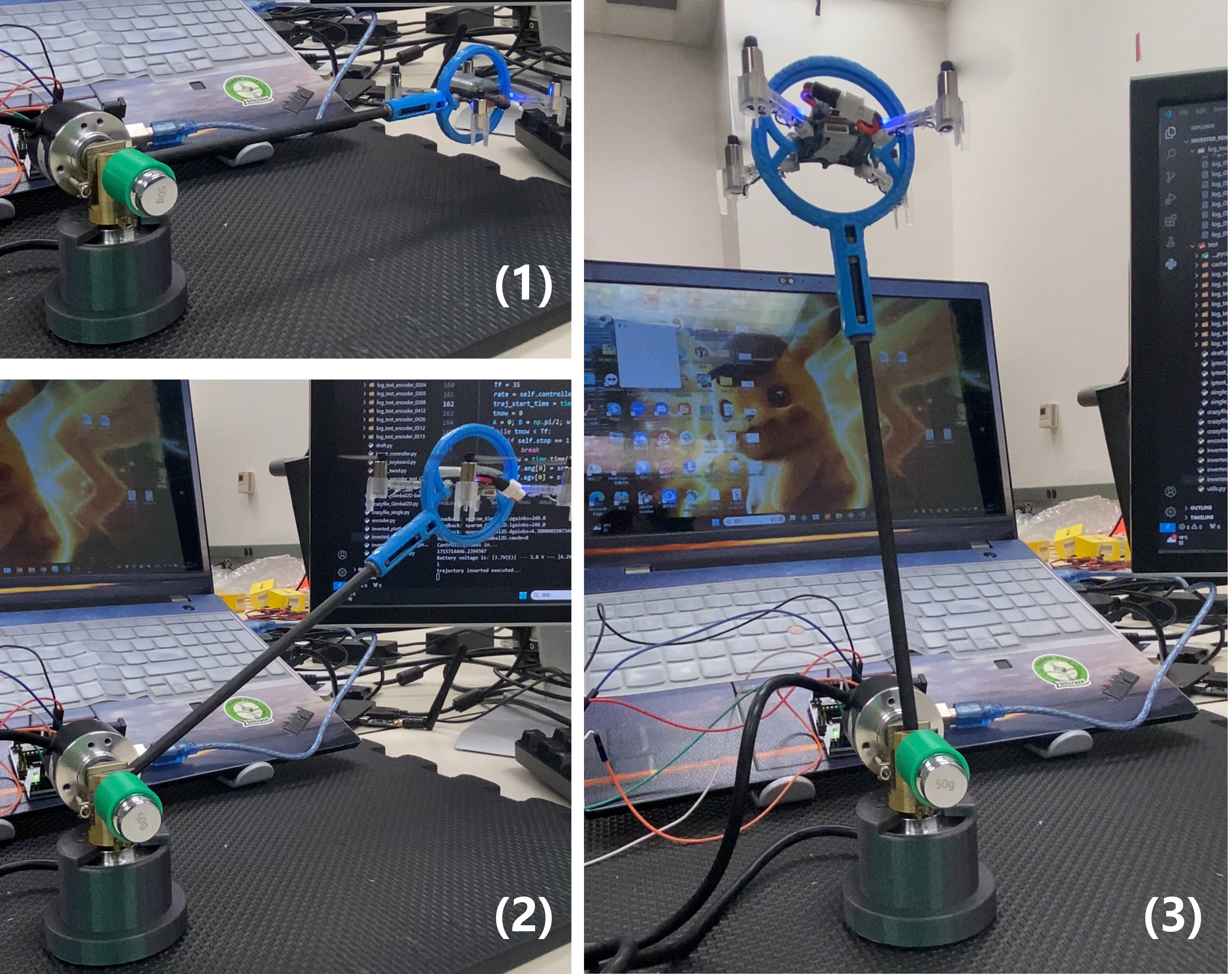}
     \caption{Inverted Trajectory Process $(1)\rightarrow(2)\rightarrow(3)$:\\$\theta_{1\_ref}=0\rightarrow\pi/4\rightarrow\pi/2$\\$\theta_{2\_ref}=0\rightarrow0\rightarrow0$}
     \label{high_level_exp_inverted_ELA_NSF_PSL_setup}
     \end{subfigure}
     \begin{subfigure}[b]{0.23\textwidth}
         \centering
         \includegraphics[width=\textwidth]{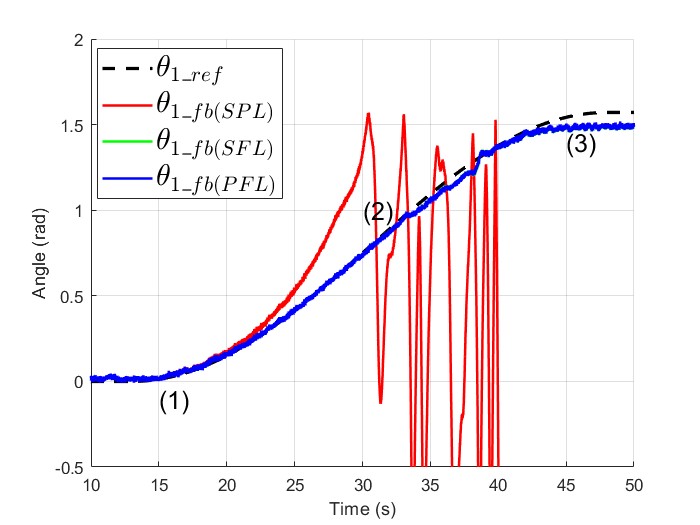}
         \caption{$\theta_1$ Simulation}
     \end{subfigure}
     \begin{subfigure}[b]{0.23\textwidth}
          \centering
          \includegraphics[width=\textwidth]{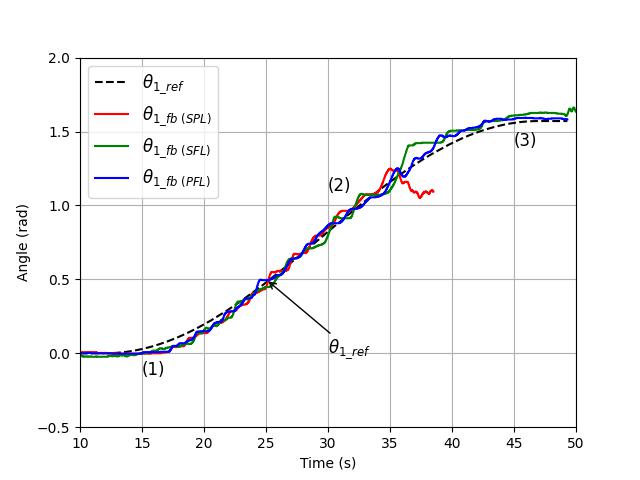}
          \caption{$\theta_1$ Experiment}
          \label{high_level_exp_inverted_theta1}
     \end{subfigure}     
     \begin{subfigure}[b]{0.23\textwidth}
         \centering
         \includegraphics[width=\textwidth]{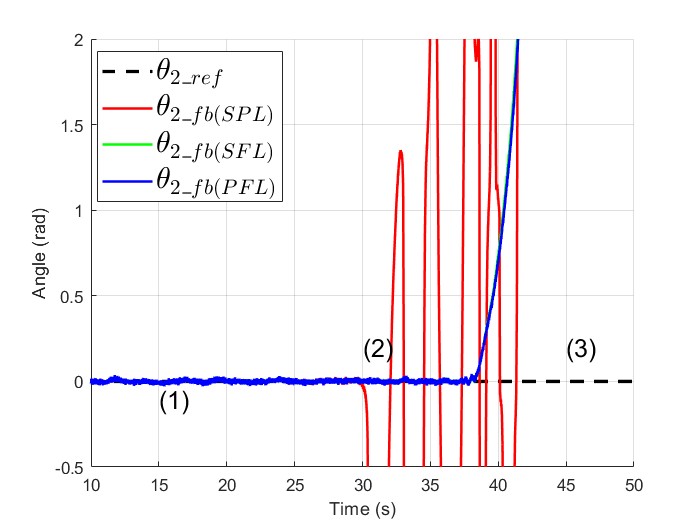}
         \caption{$\theta_2$ Simulation}
     \end{subfigure}
     \begin{subfigure}[b]{0.23\textwidth}
         \centering
         \includegraphics[width=\textwidth]{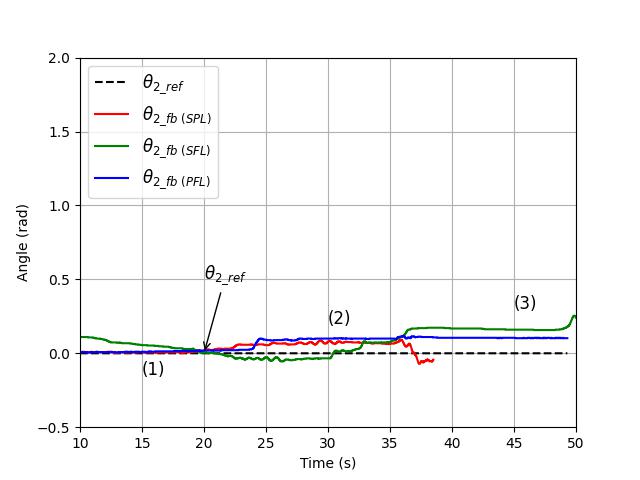}
         \caption{$\theta_2$ Experiment}
         \label{high_level_exp_inverted_theta2}
     \end{subfigure}
        \caption{Tracking towards the inverted position: $\theta_2$ is commanded to remain at $0$ while $\theta_1$ rises from $0$ to $\pi/2$ between 10 and 50 seconds with different controllers}
        \label{high_level_inverted}
\end{figure}

Tracking of a trajectory reaching the inverted position is considered in this case. The weighting matrices and the transition are the same as the previous Section \ref{V_B}. The linear controller can hardly follow the trajectory when the angle $\theta_1>0.2$ since SPL controller is linearized at $\theta_{1\_eq}=0$ and the controller performance degrades as $\theta_1$ moves away from the equilibrium position. However, the nonlinear methods achieve a consistent performance for a wide range of $\theta_1$, where the proposed platform can approach $\theta_1 = \pi/2$ and stay there with stable $\theta_2$ behavior (Fig. \ref{high_level_inverted}). 

When the pendulum crosses the singularity boundary defined in Section \ref{III_D}, a drop in the thrust magnitude evaluates the system robustness where the allocation mode is switched as discussed in Section \ref{III_C}. Therefore, a transition area is added and it solves the problem by blending control inputs between two modes, i.e. Eqs. (\ref{Fx_normal}) and (\ref{Fx_singular}). 

It is shown that SPL controller fails at $t=38$ as it leaves the equilibrium position. SFL and PFL methods are always valid while maintaining stability in the singularity region. As mentioned in Section \ref{III_D}, $\theta_2$ becomes uncontrollable when the pendulum hits the inverted position. The $T_Z$ channel is turned off to avoid infinite inputs, so $\theta_2$ might deviate from $0$ with no input capable of driving the state back to its original position. In addition, PFL controller achieves a better performance with less RMS error as indicated in Table \ref{track_towards_inv}. Even if the rotor commands saturate, the system remains stable due to the high bandwidth of the low-level controller.
\begin{table}[width=\linewidth,pos=h]
\caption{RMS Error of Inverted Tracking Performance ($\theta_1$ focused)}
\centering
\begin{tabularx}{\linewidth}{|>{\centering\arraybackslash}X|C|C|C|}
\hline
\diagbox{Case}{Unit (rad)}{Controller} & \makecell[c]{SPL \\ ($\theta_{1\_eq}=0$)} & SFL & PFL \\
\hline
Simulation & Unstable & 0.034 & 0.026 \\
\hline
Experiment & Unstable & 0.046 & 0.031 \\
\hline
\end{tabularx}
\label{track_towards_inv}
\end{table}

\section{Conclusions}\label{VI}



This paper presents a 2-DoF pendulum actuated by a 3-DoF vectored-thrust using a quadcopter, which forms a new highly-coupled nonlinear system for control problem investigation. The platform's modular architecture supports multiple control configurations, making it adaptable for varying DoF, control allocations, and nonlinear effects. The system exhibits nonlinear kinematics from the gimbals, nonlinear dynamics from the pendulum, and multi-DoF actuation from the quadcopter, making it an ideal candidate for studying both classical and modern control strategies.

We demonstrate the effectiveness of model-based control techniques, including small perturbation linearization (SPL), state feedback linearization (SFL), and partial feedback linearization (PFL), for both low-level and high-level control tasks. These methods are validated through simulation and physical experiments, showing strong agreement and reinforcing the reliability of the virtual plant model used in the control development. This confirms the practicality of the proposed modeling–simulation–implementation pipeline for nonlinear systems analysis.

Built from low-cost, off-the-shelf components and 3D-printed parts, the combined pendulum-copter platform offers a highly accessible and reproducible testbed. As such, it serves not only as a benchmark for academic research in nonlinear control, but also as a valuable educational tool for mechatronics, robotics, and control engineering curricula. The approaches hold potential for broader application in industry-oriented prototyping, testing, and deployment of UAV-based manipulation or stabilization systems. Future work includes establishing a mothership-like platform where multiple pendulum-copters are mounted to the central joint as configurable modules.

\printcredits

\appendix
\section*{Declaration of competing interest}
The authors declare that they have no known competing financial interests or personal relationships that could have appeared to influence the work reported in this paper.

\bibliographystyle{cas-model2-names}

\bibliography{ucla-refs}

\end{document}